\documentclass[aps,prb,reprint,superscriptaddress]{revtex4-1}

\usepackage[normalem]{ulem}
\usepackage{amsmath}
\DeclareMathOperator{\Tr}{Tr}
\usepackage{graphicx}
\usepackage{bm}
\usepackage{dsfont}
\usepackage{xfrac}
\usepackage{siunitx}
\usepackage{xcolor} 

\newcommand{\dg}{^{\dagger }}
\newcommand{\rarrow}{\rightarrow}
\newcommand \bea {\begin{eqnarray} }
\newcommand \eea {\end{eqnarray}}

\newcommand{\vect}[1]{\boldsymbol{\mathbf{#1}}}
\begin{document}


\title{Field-induced Ferrohastatic Order in Cubic Non-Kramers Doublet Systems}


\author{John S. Van Dyke}
\email[]{jvandyke@vt.edu}
\altaffiliation{Present address: Department of Physics, Virginia Tech, Blacksburg, Virginia 24061, USA}
\author{Guanghua Zhang}
\author{Rebecca Flint}
\affiliation{Department of Physics and Astronomy, Iowa State University, 12 Physics Hall, Ames, Iowa 50011, USA}

\date{\today}

\begin{abstract}
Cubic Pr-based compounds with $\Gamma_3$ non-Kramers doublet ground states can realize a novel heavy Fermi liquid with spinorial hybridization (`hastatic' order) that breaks time reversal symmetry.  Several Pr-``1-2-20'' materials exhibit a suggestive heavy Fermi liquid stabilized in intermediate magnetic fields; these provide key insight into the quadrupolar Kondo lattice. We develop a simple, yet realistic microscopic model of ferrohastatic order, and elaborate its experimental signatures and behavior in field, where it is a good candidate to explain the observed heavy Fermi liquids at intermediate fields in Pr(Ir,Rh)$_2$Zn$_{20}$.  In addition, we develop the Landau theory of ferrohastatic order, which allows us to understand its behavior close to the transition and explore thermodynamic signatures from magnetic susceptibility to thermal expansion.
\end{abstract}


\maketitle

\section{Introduction}


The complex interplay of spin and orbital degrees of freedom underlies many unusual properties of correlated electron systems. This interplay is especially relevant in heavy fermion materials, where it leads to exotic phenomena from unconventional superconductivity\cite{Stewart2017} and quantum criticality \cite{Coleman2005,Gegenwart2008} to  topological insulators\cite{Dzero2016}, spin liquids\cite{Lucas2017} and hidden orders\cite{Mydosh2014}. Heavy fermion research has mostly focused on Ce- or Yb-based compounds, where the $4f$ orbital is singly occupied, and its interaction with conduction electrons well described by the single-channel Kondo effect.  However, there are also Pr and U-based heavy fermion materials that contain two localized $f$ electrons and whose many-body ground state is a non-Kramers doublet protected by crystal, not time-reversal, symmetry.  These materials offer up a whole new host of behaviors driven by quadrupolar degrees of freedom and the \emph{two-channel Kondo effect}.  While these systems may form magnetic or quadrupolar order, become superconducting \cite{Flint2008,Hoshino2013} or realize a non-Fermi liquid \cite{Cox1987,Cox1998}, their Kondo physics is particularly interetsting. Here the doubly-occupied ground state fluctuates to a singly (or triply) occupied excited state.  As the excited state is Kramers degenerate, there are two distinct channels in which valence fluctuations may occur. Heavy fermions may still form, but now require breaking the channel symmetry.  This symmetry-broken heavy Fermi liquid has been termed either ``diagonal composite order'' \cite{Hoshino2011,Hoshino2013,Kuramoto2014} or, to emphasize its novel spinorial nature, ``hastatic order'' \cite{Chandra2013,Chandra2015,Zhang2018}.  Hastatic order is a \emph{fractionalized} order \cite{Komijani2018}, with a spinorial hybridization. 

Cubic materials provide a particularly simple setting in which to study this physics, as here the two-channel Kondo effect is a Kondo effect for the local quadrupolar moments, which are screened by conduction quadrupolar moments in two different spin channels. Indeed, the physics of the quadrupolar Kondo lattice is a long standing problem. In particular, its two-channel nature and relevance to the non-Fermi liquid and unconventional superconductivity in UBe$_{13}$ \cite{Cox1987,Cox1998} are not fully understood.  However, discerning its role in actinide materials is challenging due to difficulties in resolving the valence and crystal field ground states. The recently discovered cubic Pr-based 1-2-20 materials provide an important opportunity to study these phenomena in a simpler system.  These materials exhibit Kondo physics at high temperatures \cite{Sakai2011,Onimaru2011,Matsunami2011,Onimaru2012,Tsujimoto2014,Shimura2015,Onimaru2016a}, along with quadrupolar \cite{Sakai2011,Onimaru2011,Ito2011,Onimaru2012,Sato2012,Ishii2013,Taniguchi2016,Iwasa2017}, superconducting \cite{Onimaru2010,Sakai2012,Matsubayashi2012,Onimaru2012,Tsujimoto2014}, non-Fermi liquid \cite{Matsubayashi2012,Shimura2015} and unidentified low temperature phases \cite{Tsujimoto2014,Tsujimoto2015,Onimaru2016b,Yoshida2017}.  Unlike in the actinides, the ground state is known to be the $4f^2$ $\Gamma_3$ \cite{Sakai2011,Onimaru2011,Onimaru2012}, which imposes two-channel Kondo physics. These materials provide an ideal setting to resolve the role of the quadrupolar Kondo effect and explore hastatic order within a simpler setting. Several exhibit a dome of heavy Fermi liquid at finite magnetic fields \cite{Onimaru2016b,Yoshida2017} that is consistent with field-induced hastatic order.  

While our previous work has explored cubic hastatic order in a simple two-channel Kondo model \cite{Zhang2018}, comparison to experiment requires a more realistic model.  To this end, we treat uniform or ``ferrohastatic'' (FH) order in the realistic cubic two-channel Anderson model with a combination of a microscopically motivated mean-field theory (justified within large-$N$ and expected to work well at low temperatures) and a phenomenological Landau theory (which can capture the nature of the phase transition).  Neither approach captures the whole story, but together they give significant insight.  Finally, we argue that the intermediate field regions in Pr(Ir,Rh)$_2$Zn$_{20}$ are ferrohastatic, and give concrete experimental tests, including induced dipole moments in magnetic field, signatures in magnetostriction and thermal expansion, spin-resolved spectroscopies, and novel symmetry-breaking hybridization gaps.

\subsection{Pr-based 1-2-20 materials details}

Kondo physics in Pr-based materials is rare, but the 1-2-20 materials, Pr$T_2X_{20}$, have atypically strong $c$--$f$ hybridization \cite{Sakai2011,Tokunaga2013}, as the Pr ions sit within Frank-Kasper cages of 16 X=Al or Zn atoms. Cubic crystal fields select a $\Gamma_3$ ground state doublet \cite{Sakai2011,Onimaru2011,Onimaru2012}, and there is considerable evidence for Kondo physics: at high temperatures, experiment shows partial quenching of the $R\ln 2$ entropy \cite{Sakai2011}, logarithmic scattering terms in the resistivity \cite{Tsujimoto2014}, large hyperfine coupling due to $c$--$f$ hybridization \cite{Tokunaga2013}, enhanced effective masses \cite{Shimura2015}, and a Kondo resonance in photoemission \cite{Matsunami2011}. At low temperatures, all of these materials order in some fashion and become superconducting at very low temperatures: PrTi$_2$Al$_{20}$ and PrIr$_2$Zn$_{20}$ order ferro- (FQ) and antiferroquadrupolarly (AFQ) at $T_Q = 2$K and $0.11$K, respectively \cite{Sakai2011,Onimaru2011}, while the ordering in PrV$_2$Al$_{20}$ \cite{Sakai2011} and PrRh$_2$Zn$_{20}$ \cite{Onimaru2012} is still undetermined, although octupolar order seems likely in PrV$_2$Al$_{20}$ \cite{Freyer2018,Lee2018,Patri2018}. Quadrupolar order can be suppressed both with pressure (PrTi$_2$Al$_{20}$ \cite{Matsubayashi2012}) and with field [Pr(Ir,Rh)$_2$Zn$_{20}$ \cite{Onimaru2011,Onimaru2012} and PrV$_2$Al$_{20}$ \cite{Sakai2011}], leading to an extended non-Fermi liquid region at higher temperatures. Pressure enhances the superconductivity in PrTi$_2$Al$_{20}$ \cite{Matsubayashi2012}, which is likely unconventional.  The in-field phase diagrams are even more interesting, as there is a heavy Fermi liquid region sandwiched between the zero-field order and a polarized high field state where Kondo physics is lost\cite{Onimaru2016b,Yoshida2017}. 

\subsection{Structure of the paper}

Ferrohastatic order and the generic infinite-$U$ two-channel Anderson model is introduced in section \ref{sec:FHorder}.  Section \ref{sec:microscopics} fleshes out the details of the microscopic Anderson model and solves it within a large-$N$ mean-field theory for both FH and the competing antiferrohastatic (AFH) orders, and also considers interactions with the competing antiferroquadrupolar (AFQ) order.  In section \ref{sec:landau}, we develop the Landau theory of cubic ferrohastatic order, examine its interactions with field, strain and AFQ order and discuss the thermodynamic signatures. Finally, in Section \ref{sec:signatures} we summarize and expand upon the experimental signatures of FH order and how it may be distinguished from quadrupolar orders, before concluding in Section \ref{sec:conclusion}.

\section{Ferrohastatic order \label{sec:FHorder}}

\begin{figure}[htbp]
\vspace*{-0cm}
 \includegraphics[scale=0.45]{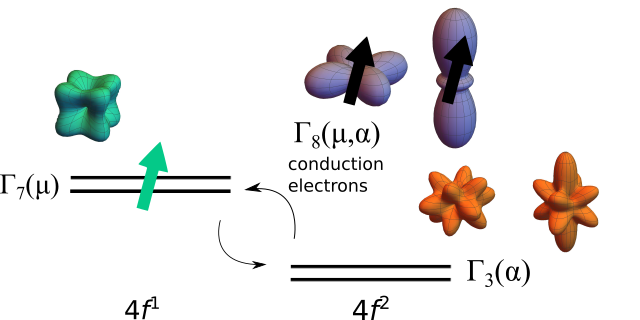}%
\vskip -0.2cm
 \caption{\small Atomic level diagram illustrating valence fluctuations out of a $4f^2$ non-Kramers ($\Gamma_3$) doublet to a $4f^1$ excited $\Gamma_7$ Kramers doublet via the conduction electron quartet, $\Gamma_8$. The form-factors of each state are shown, with the $\Gamma_7$ and $\Gamma_8$ orbitals possessing an extra dipolar moment (indicated by the arrows).\label{fig:leveldiagram}}
 \end{figure}

Hastatic order is a natural candidate for materials with an even number of $f$ electrons and doublet crystal-field ground states. The cubic $\Gamma_3$ doublet is the simplest of these, with no dipole moments, $\langle \vec{J} \rangle = 0$, but finite quadrupolar ($O_{x^2-y^2}$, $O_{3z^2-r^2}$) and octupolar ($T_{xyz}$) moments \cite{Cox1998}, and it is protected by cubic, not time-reversal, symmetry.  Overlap between the non-Kramers $\Gamma_3$ states and conduction electrons leads to valence fluctuations, shown in Fig. \ref{fig:leveldiagram}, in which a $4f$ electron escapes into the conduction sea, leaving an excited $4f^1$ state, here the $\Gamma_7$ Kramers doublet \footnote{The transition to $4f^3$ is more likely in the real materials, but yields the same physics as the $4f^1$ transition considered here}. These valence fluctuations are mediated by a $\Gamma_8$ quartet of conduction electrons, and thus two conduction channels screen a single $f$-moment. We consider a simple cubic lattice with two $e_g$ conduction bands that have the required $\Gamma_8=e_g\otimes\sfrac{1}{2}$  symmetry, from the orbital and spin degrees of freedom, respectively \footnote{The 1-2-20 materials have an fcc structure, but the nature of FH order is independent of the structural details.}.  The $\Gamma_3$ states are labeled by their quadrupole moments, $\alpha$, while the excited $\Gamma_7$ are labeled by their dipole moments, $\mu$. $\mu$ is the channel index in a two channel Anderson lattice model \cite{Cox1998}, while $\alpha$ represents the screened pseudospin.

We consider an infinite-$U$ two channel Anderson model with the Hamiltonian:
\begin{align}
\mathcal{H} &= \mathcal{H}_c + \mathcal{H}_f + \mathcal{H}_{VF}.
\end{align}
The valence fluctuation Hamiltonian is
\begin{align}
\mathcal{H}_{VF} = V \sum_{j\mu\alpha}  \tilde{\mu} |j, \Gamma_3, \alpha \rangle \langle j, \Gamma_7, -\mspace{2mu} \mu | \psi_{j, \Gamma_8 \mu \alpha} + H.c. \label{eq:VF}
\end{align}
The Hubbard operators $| j,\Gamma_3,\alpha\rangle \langle j, \Gamma_7,-\mspace{2mu}\mu |$ transition the $f$-electron system between the ground and excited states, while $\psi_{j,\Gamma_8 \mu \alpha}$ annihilates a $\Gamma_8$ conduction electron. $V$ is the bare hybridization strength and $\tilde{\mu} = \mathrm{sgn}(\mu)$ imposes a singlet state of the conduction and $f$ electrons.  The conduction and $f$-electron terms are
\begin{align}
\mathcal{H}_c &= \sum_{\vect{k}\sigma\alpha\beta}  \epsilon_{\vect{k}\alpha\beta} c^\dagger_{\vect{k}\sigma\alpha} c_{\vect{k}\sigma\beta}   \label{eq:Hc} \\
 \mathcal{H}_f &= \sum_{j\mu}  \Delta E  | j,\Gamma_7,\mu\rangle \langle j, \Gamma_7,\mu |, \label{eq:Hf}
\end{align}
where $c_{\vect{k}\sigma\alpha}$ annihilates a conduction electron in channel $\sigma$ with pseudospin $\alpha=\{+,-\}$ ($\epsilon_{\vect{k}\alpha\beta}$ is the conduction electron dispersion).  Here $\Delta E > 0$ is the energy of the excited $4f^1$ state and $ | j,\Gamma_7,\mu\rangle \langle j, \Gamma_7,\mu |$ is the projector onto this state.

To proceed, we replace the Hubbard operators with slave bosons $b_{j\mu}$ and fermions $f_{j\alpha}$ \cite{Coleman1984,Read1983}. $b_{j\mu}$ represents the excited doublet and $f_{j\alpha}$ the non-Kramers doublet.  Other states are forbidden, imposed by the constraint $f^\dagger_{j\alpha} f_{j\alpha} + b_{j\mu}^\dagger b_{j\mu}  = 1$, where we introduce Einstein summation notation. The Hubbard operators become
\begin{align}
& |j,\Gamma_7,\mu\rangle \langle j,\Gamma_7,\mu | \rightarrow b^\dagger_{j\mu} b_{j\mu}, \cr
&|j,\Gamma_3,\alpha\rangle \langle j,\Gamma_3,\alpha | \rightarrow f^\dagger_{j\alpha} f_{j\alpha},\cr
& |j,\Gamma_3,\alpha\rangle \langle j,\Gamma_7,\mu | \rightarrow f^\dagger_{j\alpha} b_{j\mu},
\end{align}
In this representation, the Hamiltonian becomes
\begin{align}
H = & \sum_{\vect{k}\sigma} \epsilon_{\vect{k}\alpha\beta} c^\dagger_{\vect{k}\sigma\alpha} c_{\vect{k}\sigma\beta}  +\! V\! \sum_{j}\! \left( \tilde{\mu} f^\dagger_{j\alpha} b_{j-\mu} \psi_{j,\Gamma_8 \mu \alpha} + H.c. \right)\notag \\
&  + \sum_j \left( \left[\lambda_j + \Delta E \right] b_{j\mu}^\dagger b_{j\mu} + \lambda_j  \left[f^\dagger_{j\alpha} f_{j\alpha} - 1\right] \right), 
\end{align}
where the Lagrange multipliers $\lambda_j$ enforce the constraint. 

\begin{figure}[htbp]
\vspace*{-0cm}
 \includegraphics[scale=0.60]{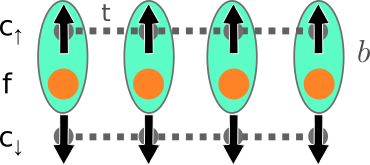}%
\vskip -0.2cm
 \caption{\small Schematic of ferrohastatic (FH) order in which the $f$-electron hybridizes exclusively with the spin-$\uparrow$ conduction electrons. $b$ represents the occupation of the excited state, or in the Kondo limit, the Kondo singlet formed between local moment and conduction electrons.\label{fig:FHansatz}}
 \end{figure}

This model can be solved exactly within an $SU(N)$ large-$N$ limit, where $\alpha = \pm 1, \ldots ,N$, while $\mu = \uparrow,\downarrow$ remains $SU(2)$.  In this mean-field limit, the slave bosons condense, $\langle b_{j\mu} \rangle \neq 0$ below the transition temperature $T_K$. On account of the two degenerate excited levels (corresponding to the channels labeled by $\mu$), the hastatic order parameter $b$ forms a spinor
\begin{align}
b = \begin{pmatrix}
b_\uparrow \\
b_\downarrow
\end{pmatrix} \label{eq:bspinor}
\end{align}
As $b_\uparrow$ and $b_\downarrow$ assume definite values in the hastatic state, the system necessarily breaks time reversal and spin rotation symmetry.  While these are also broken in an ordinary magnetic system, hastatic order additionally breaks double time reversal symmetry, due to the spinorial nature of the order parameter.  Microscopically, we may think of hastatic order as consisting in a choice of hybridization spinor (magnitude and direction) for each site in the lattice.  This leads to various realizations of hastatic order similar to the forms of magnetic order (ferro-, antiferro-, etc.) determined by the arrangement of spins in a magnetic system.   We term the simplest possibility, namely a uniform magnitude and direction of the spinor at each site, \textit{ferrohastatic order} (FH), in analogy with the magnetic case.  A particular FH ansatz, in which the $f$ electrons exclusively hybridize with spin-$\uparrow$ conduction electrons, is shown in Fig. \ref{fig:FHansatz}.

The large-$N$ phase diagram of the two-channel Kondo limit shows that FH order is favored in a range around half-filling, with antiferrohastatic (AFH) order favored for smaller fillings \cite{Zhang2018}.  Strong coupling analysis yields a similar picture, where the strong coupling limit of our model is the two-channel Kondo lattice with $J_K = \frac{V^2}{\Delta E}$.  As $J_K \rightarrow \infty$, conduction electrons added to the system form Kondo singlets until all of the local moments are screened, which occurs at quarter-filling.  These Kondo singlets carry the channel (physical spin) index and can be treated as hard-core bosons \cite{Schauerte2005}.  Exactly at quarter-filling, these spinful Kondo singlets are the only degree of freedom and order antiferrohastatically due to superexchange ($\sim t^2/J_K$) from virtual hopping ($t$) of the conduction electrons.  Adding a single conduction electron forces the Kondo singlets to be FH in order to maximize kinetic energy ($\sim t$), in analogy with the infinite-U Hubbard model \cite{Nagaoka1966}; as in the Hubbard model, we expect the AFH region to extend some distance above quarter-filling for finite $J_K$.  FH order also wins at half-filling, as it again maximizes the kinetic energy.  Note that hastatic order is always stabilized over quadrupolar order at strong coupling, as the local Kondo singlet lowers its energy via quantum fluctuations, while AFQ order freezes the local $f$-moment.  On site, the energy of the Kondo singlet is $-J_K S(S+1) = -3J_K/4$, while the frozen $f$-moment minimizes its energy with two conduction electrons per site: $ c^\dagger_{\uparrow +} f^\dagger_- c^\dagger_{\downarrow +}\vert 0\rangle$, with energy $-2 J_K S^2 = -J_K/2$.
In section \ref{sec:microscopics}, we solve our Anderson model within the $SU(N)$ large-$N$ limit and find that FH order is found in a large region around half-filling of the conduction electrons, similar to what is expected from this strong coupling analysis and what was found in the Kondo limit. Additionally, we shall see that as FH order contains small magnetic moments, it is favored by magnetic field, as is also the case in the Kondo limit \cite{Zhang2018}.

\section{Microscopic model and phase diagrams \label{sec:microscopics}}

Now we return to our microscopic Anderson model to flesh out the details and solve it within the large-$N$ limit.  As we are particularly interested in the effect of magnetic field, we first examine this coupling in detail.

Magnetic field affects Kramers and non-Kramers components differently, coupling linearly to the conduction electrons and the excited $\Gamma_7$ state, as shown in the Hamiltonian below, where the magnetic field $\vec{B} = B \hat z$.  $\Gamma_3$ does not couple to $B$  directly, but acquires a small moment linear in $B$ due to virtual transitions to the excited triplet states at energy $\Delta$, leading to an $O(B^2/\Delta)$ splitting \cite{Zhang2018}.  For simplicity, we consider transitions only to the excited $\Gamma_4$ triplet \cite{Onimaru2016a}, which affects only $\vert\Gamma_3, +\rangle$.  The magnetic field part of the Hamiltonian is therefore 
\begin{align}
H_B = &-\sum_{\vect{k}\sigma\alpha\beta} \tilde{\sigma} \mu_B B  c^\dagger_{\vect{k}\sigma\alpha} c_{\vect{k}\sigma\beta} -\sum_j \Big( \gamma B^2 \delta_{\alpha,+} f^\dagger_{j\alpha} f_{j\alpha} \notag \\
& \hspace{20pt}  +  \mu_B g_L B \langle J_z \rangle_{\Gamma_7} \tilde{\mu} b_{j\mu}^\dagger b_{j\mu}  \Big).
\end{align}
$\mu_B$ is the Bohr magneton, $g_L$ the Land\'{e} $g$-factor and $\langle J_z \rangle_{\Gamma_7}$ the $J_z$ angular momentum of the $\Gamma_7$ state.  $\gamma = 6$ gives the nonlinear coupling of $|\Gamma_3,+\rangle$ to $B^2$.  In finite field, the $\vert\Gamma_3, +\rangle$ state develops a dipole moment linear in field, 
\begin{equation}
\!m_f = \mu_B \!\! \left[ 12 \frac{\mu_B B}{\Delta}\! -\! 130 \left(\frac{\mu_B B}{\Delta}\right)^3\!\right]\! \langle n_{g} \rangle + O[\left(\frac{\mu_B B}{\Delta}\right)^5\!],\!
\end{equation}
for $\mu_B B \ll \Delta$, where $\langle n_{g} \rangle$ is the ground state ($\Gamma_3+$) occupation.

The slave boson Hamiltonian is then,
\begin{align}
H = & \sum_{\vect{k}\sigma} \left( \epsilon_{\vect{k}\alpha\beta} -\mu_B B \tilde{\sigma} \right) c^\dagger_{\vect{k}\sigma\alpha} c_{\vect{k}\sigma\beta} \notag \\
& + V \sum_{j} \left( \tilde{\mu} f^\dagger_{j\alpha} b_{j-\mu} \psi_{j,\Gamma_8 \mu \alpha} + H.c. \right)\notag \\
&  + \sum_j \left( \left[\lambda_j + \Delta E -\mu_B B g_L \langle J_z \rangle_{\Gamma_7} \tilde{\mu} \right] b_{j\mu}^\dagger b_{j\mu} \right. \notag \\
& \hspace{20pt} + \left. \left[ \lambda_j - \gamma B^2 \delta_{\alpha,+} \right]  f^\dagger_{j\alpha} f_{j\alpha} - \lambda_j \right), 
\end{align}

The $\Gamma_8$-symmetry electrons, $\psi_{j,\Gamma_8 \mu \alpha}$ appearing above are $f$-states, but they have a finite overlap with the conduction electron bands of whatever type, which can be incorporated via a Wannier form factor $\Phi$; here this describes the overlap between the odd-parity $\Gamma_8$ and the even-parity $d$ states at neighboring sites:
\begin{align}
\psi_{j, \Gamma_8 \mu \alpha} &= \sum_{\vect{k} \sigma \alpha'} \mathrm{e}^{i \vect{k} \cdot \vect{R}_j} \Phi^{\sigma \alpha'}_{\mu \alpha}(\vect{k})
c_{\vect{k} \sigma \alpha'}.
\end{align}
For simplicity we consider Pr $5d$ states; $f$-electrons in PrT$_2$(Al,Zn)$_{20}$ are more likely to hybridize with Al or Zn $p$-states, leading to different form factors but qualitatively similar physics.  Our bands mirror those of SmB$_6$, where the $\Gamma_8$ ground state couples to $e_g$ conduction electrons \cite{Alexandrov2013,Baruselli2014}, although the nature of our (spinorial) hybridization is clearly different. We consider generic nearest-neighbor conduction electron dispersions $\epsilon_{\vect{k}}$ and hybridization form factors $\Phi$ with cubic symmetry. Both of these are matrices: $\epsilon_{\vect{k}}$ is a matrix in $\alpha$ and $\sigma$ space, and is derived similarly to the hybridization form factor, shown in Appendix \ref{sec:slaterkoster},
 \begin{align}
 \mathcal\epsilon_{\vect{k}} = &
  - t[(c_x+c_y)(\frac{1}{2} + \frac{3}{2}\eta_c)+2c_z] (\sigma_0 \otimes\alpha_0) \cr 
  &  -\frac{\sqrt{3}}{2}t(c_x-c_y)(1-\eta_c)(\sigma_0 \otimes\alpha_1)\cr
  & -\mu_B B (\sigma_3 \otimes \alpha_0) \label{eq:Hd}
\end{align}
where $c_i \equiv \cos (k_i a)$ $(i=x,y,z)$ and $\mu$ is the chemical potential. For our numerical calculations we set the nearest neighbor spacing $a = 1$ and the overall hopping magnitude $t=1$, effectively measuring everything else in units of $t$.  The conduction electron band width $D = 12t$.  There is a single free parameter, $\eta_c$ that tunes the degeneracies and anisotropies of the bands. We fix the number of conduction electrons above the transition, $n_{c0}$ and allow $\mu$ to vary to preserve the total charge. The hybridization form factors $\Phi$ are given below, but are similarly described by an overall magnitude $V$ and free parameter $\eta_V$.

\subsection{Slave boson theory for ferrohastatic order}

In this section, we give the full detailed mean-field Hamiltonian for the FH ansatz with the hastatic spinor oriented along $\hat{z}$, $\hat{b}_j = (b, 0)^T$,
\begin{align}
H & = \sum_{\vect{k}} \epsilon_{\vect{k}\alpha\beta} c^\dagger_{\vect{k}\sigma\alpha} c_{\vect{k}\sigma\beta} + \mathcal{N}\left(\Delta E -\mu_B B g_L \langle J_z \rangle_{\Gamma_7}\right)|b|^2  \cr
&\!\! - Vb\! \sum_{j}\! \left(  f^\dagger_{j\alpha} \psi_{j,\Gamma_8 \downarrow \alpha} + H.c. \right)\! +\! \lambda \sum_{j} (f^\dagger_{j\alpha} f_{j\alpha} +  |b|^2\!  -\! 1)  \notag \\
&
\!\!+  \mu \sum_j(c^\dagger_{j\sigma\alpha} c_{j\sigma\alpha}\! -\! |b|^2\! -\! n_{c,0})\!-\! \gamma B^2 \sum_{j}\! f^\dagger_{j+} f_{j+} 
\end{align}
Here, $\epsilon_{\vect{k}\alpha\beta}$ is the conduction electron dispersion matrix given in eqn. (\ref{eq:Hd}).  There are two Lagrange multipliers, $\lambda$ and $\mu$.  The first enforces the average local constraint on the occupations of the localized $f$-electron orbitals, while the second enforces the global conservation of charge.  The magnetic field lies solely along the direction of the hastatic spinor, $\hat{z}$.  In momentum space, the Hamiltonian is
\begin{align}
H &= \sum_{\vect{k},\sigma} \epsilon_{\vect{k}\alpha\beta} c^\dagger_{\vect{k}\sigma\alpha} c_{\vect{k}\sigma\beta} + \mathcal{N}\Delta E |b|^2 + \lambda \mathcal{N} (|b|^2 - 1) \notag \\
&- Vb \sum_{\vect{k}\sigma\alpha\alpha'} \Big(  f^\dagger_{\vect{k},\alpha} c_{\vect{k}, \sigma \alpha'}\Phi^{\sigma \alpha'}_{\downarrow \alpha}(\vect{k}) 
+ H.c. \Big) \notag \\
&+ \mu \left[ \sum_{\vect{k}} c^\dagger_{\vect{k}\sigma\alpha} c_{\vect{k}\sigma\alpha} - \mathcal{N} (|b|^2 - n_{c,0}) \right]+\lambda \sum_{\vect{k}}f^\dagger_{\vect{k}\alpha}f_{\vect{k}\alpha} \notag \\
&- \! \gamma B^2\! \sum_{\vect{k}}  f^\dagger_{\vect{k}+} f_{\vect{k}+}  -  \mathcal{N} g_L \langle J_z \rangle_{\Gamma_7}  \mu_B B |b|^2 
\end{align}
where $\mathcal{N}$ is the number of sites.  In a path integral approach, the saddle-point approximation (exact in the $SU(N)$  large-$N$ limit) leads to the self-consistency equations:
\begin{align}
\frac{\partial \mathcal{F}}{\partial b} &= 0;  &
\frac{\partial \mathcal{F}}{\partial \lambda} &= 0; & 
\frac{\partial \mathcal{F}}{\partial \mu} &= 0.  &
\label{eq:selfcons}
\end{align}
The resulting mean field Hamiltonian can be written as a matrix
\begin{align}
H &= \sum_{\vect{k}} \Psi_{\vect{k}}^\dagger \begin{pmatrix}
\mathcal{H}_c(\vect{k}) & \mathcal{V}_z(\vect{k})^\dagger \\
\mathcal{V}_z(\vect{k}) & \mathcal{H}_f(\vect{k})
\end{pmatrix}\Psi_{\vect{k}}  + const. \notag \\
&\equiv \sum_{\vect{k}} \Psi_{\vect{k}}^\dagger \mathcal{H}_{\vect{k}}'
\Psi_{\vect{k}} + const.\label{eq:HMF}
\end{align}
with spinor $\Psi = (
 c_{\uparrow +} \; c_{\uparrow -} \; c_{\downarrow +} \; c_{\downarrow -} \; f_{+} \; f_{-})^T$.  Here $\mathcal{H}_c = \epsilon_{\vect{k}} + \mu \sigma_0 \alpha_0$ is a $4 \times 4$ matrix, $\mathcal{H}_f = \lambda \alpha_0-\gamma B^2 (1+\alpha_3)/2$ is a $2 \times 2$ matrix (we use two types of Pauli matrices, $\sigma_\lambda$ and $\alpha_\lambda$ ($\lambda = 0,1,2,3$), to represent the spin and pseudospin degrees of freedom), and $\mathcal{V}_z$ is the $2 \times 4$ hybridization matrix,
\begin{align}
\mathcal{V}(\vect{k}) = -V\sum_{\mu} \tilde{\mu} b_{-\mu} \Phi^{\sigma \alpha'}_{\mu \alpha}(\vect{k}),
\end{align}
which takes the form for $\hat b || \hat z$ ,

\begin{align}
&\mathcal{V}_z(\vect{k}) = -Vb\begin{pmatrix}
  \Phi^{\uparrow +}_{\downarrow +}(\vect{k})   &  \Phi^{\uparrow -}_{\downarrow +}(\vect{k}) & \Phi^{\downarrow +}_{\downarrow +}(\vect{k}) & \Phi^{\downarrow -}_{\downarrow +}(\vect{k}) \\
\Phi^{\uparrow +}_{\downarrow -}(\vect{k}) &  \Phi^{\uparrow -}_{\downarrow -}(\vect{k})  & \Phi^{\downarrow +}_{\downarrow -}(\vect{k})  &  \Phi^{\downarrow -}_{\downarrow -}(\vect{k})  
\end{pmatrix}\\
&= -Vb \begin{pmatrix}
-\frac{i(1\!+\!3\eta_v)}{2}s_+ & \frac{i\sqrt{3}(-1\!+\!\eta_v)}{2}s_- & -2i s_z & 0 \\
\frac{i\sqrt{3}(-1\!+\!\eta_v)}{2}s_-  & -\frac{i(3\!+\!\eta_v)}{2}s_+ & 0 & -2i \eta_v s_z
\end{pmatrix}.
\end{align}
$s_\pm = s_x \pm i s_y$, where $s_i = \sin (k_i a) (i = x,y,z)$. The free energy density $\mathcal{F}$ is obtained by integrating out the fermions, leading to
\begin{align}
\mathcal{F} =& -T\sum_{\vect{k}\zeta} \ln (1 + e^{-E_{\vect{k}\zeta}/T}) + (\lambda+\Delta E)|b|^2 - \lambda  \notag \\
&- \mu (|b|^2 + n_{c,0})- g_L \langle J_z \rangle_{\Gamma_7} \tilde{\mu} B|b|^2
\end{align}
where $E_{\vect{k}\zeta}$ are the six eigenvalues of the Hamiltonian matrix $\mathcal{H}_{\vect{k}}'$. Within this formalism, we can treat both FH order ($b \neq 0$) and paraquadrupolar order ($b = 0,\lambda =0$).  From the Kondo limit and strong coupling analysis, we expect FH order to be favored near half-filling, and as the system gains energy by aligning the spinor with the external field, the uniform FH case is also favored in field over competing states with non-uniform arrangements of the hastatic spinor. In section \ref{sec:FHvsAFQ}, we consider the competition between the FH and the AFQ ansatzes.

\begin{figure}[ht]
 \includegraphics[scale=0.5]{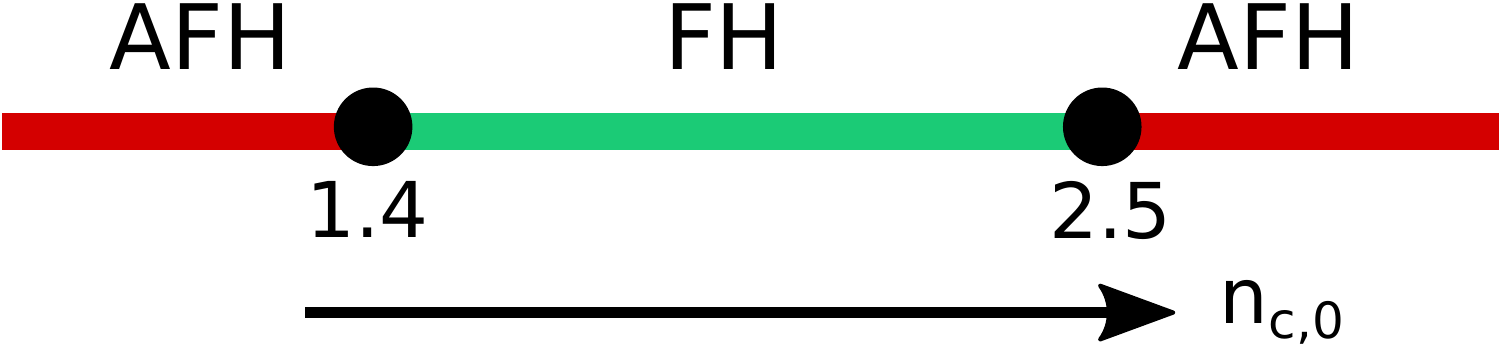}%
 \caption{Phase diagram of ferrohastatic (FH) and antiferrohastatic (AFH) orders at $T=0$ as a function of conduction electron filling.  FH order is favored near half-filling, while AFH order is favored near quarter-filling, as expected from the strong coupling approach. Note that particle-hole symmetry is absent in the Anderson model approach due to the finite occupation of the excited $f$ state.
 \label{fig:AFHPD}}
 \end{figure}
 
\subsection{Antiferrohastatic mean-field theory}

\begin{figure}[ht]
 \includegraphics[scale=0.60]{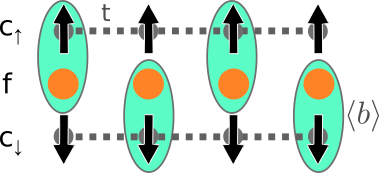}%
 \caption{Cartoon of AFH order in 1D, where the hastatic spinor $\hat{b}$ is alternately aligned with the $+\hat{z}$ and $-\hat{z}$ axes. \label{fig:AFHcartoon}}
 \end{figure}

In Fig. \ref{fig:AFHPD} we show the mean-field $T = 0$ phase diagram as a function of conduction electron filling $n_{c0}$ for our model on the simple cubic lattice.  As expected from the strong coupling analysis, FH order appears near half-filling, while AFH is found around quarter-filling.  Here we used the mean-field theory of the N\'{e}el staggered AFH ansatz to compute the phase diagram.  The two N\'{e}el sublattices have the hastatic spinor oriented oppositely, e.g. $\hat{b}_A = (b, 0)^T$, $\hat{b}_B = (0, b)^T$.  A one dimensional version of this AFH order is shown in the cartoon of Fig. \ref{fig:AFHcartoon}.

The slave boson expectation value at site $j$ may be written as
\begin{align}
b_j &=  \frac{b_A}{2} \left(1 + e^{i \vect{Q} \cdot \vect{R}}  \right) +  \frac{b_B}{2} \left(1 -  e^{i \vect{Q} \cdot \vect{R}}  \right) \notag \\
&= \frac{1}{2}\left[
\begin{pmatrix}
b\\
b
\end{pmatrix}
+e^{i \vect{Q} \cdot \vect{R}}
\begin{pmatrix}
b\\
-b
\end{pmatrix}
\right],
\end{align}
where $\vect{Q} = (\pi,\pi,\pi)$.  The mean-field Hamiltonian for the AFH case can then be written, 
\begin{align}
H =& \sum_{\vect{k},\sigma}  \epsilon_{\vect{k}\alpha\beta} c^\dagger_{\vect{k}\sigma\alpha} c_{\vect{k}\sigma\beta}  \cr
& + \frac{Vb}{2}\! \sum_{\vect{k}}\! \left( [\Phi_{\sigma \alpha'}^{\uparrow \alpha}(\vect{k}) - \Phi_{\sigma \alpha'}^{\downarrow \alpha}(\vect{k})]c^\dagger_{\vect{k} \sigma \alpha'} f_{\vect{k},\alpha} \right. \cr
& \hspace{1em} \left. - [\Phi_{\sigma \alpha'}^{\uparrow \alpha}(\vect{k}) + \Phi_{\sigma \alpha'}^{\downarrow \alpha}(\vect{k})]c^\dagger_{\vect{k} \sigma \alpha'} f_{\vect{k}+\vect{Q},\alpha} + H.c. \right) \notag \\
& +  \lambda \left[ \sum_{\vect{k}} f^\dagger_{\vect{k}\alpha} f_{\vect{k}\alpha}  +  \mathcal{N} (|b|^2 - 1)  \right] + \mu \sum_{\vect{k}} c^\dagger_{\vect{k}\sigma\alpha} c_{\vect{k}\sigma\alpha} \cr
&- \mu \mathcal{N} (|b|^2 - n_{c,0})  + \mathcal{N}\Delta E |b|^2
\end{align}
where $\vect{k}$ ranges over the original Brillouin zone.  The free energy is obtained in a similar fashion as the FH case,
\begin{align}
\mathcal{F}  =& -T\sum_{\vect{k}\zeta} \ln (1 + e^{-E_{\vect{k}\zeta}/T}) + \mathcal{N}(\lambda+\Delta E)|b|^2 \cr &- \mathcal{N}\lambda 
-\mu \mathcal{N}(|b|^2 + n_{c,0}),
\end{align}
where $E_{\vect{k}\zeta}$ now ranges over the twelve AFH bands.  The free energy is minimized by solving the saddle point equations, $\partial \mathcal{F}/\partial b = 0$, $\partial \mathcal{F}/\partial \lambda = 0$, and $\partial \mathcal{F}/\partial \mu = 0$.

The AFH phase has staggered magnetic moments, but no uniform moments (magnetic or multipolar), and we find that FH order is quickly favored over AFH in finite magnetic field.  The relative stability of AFH order will be more materials-dependent than FH order, as it depends more strongly on the details of the crystal structure.  Here we analyze the simple cubic case for simplicity, but the qualitative features are expected to hold for the diamond lattice applicable to the Pr-1-2-20 materials.

\subsection{Competition with antiferroquadrupolar order \label{sec:FHvsAFQ}}

To capture the competing AFQ orders observed in PrT$_2$X$_{20}$ materials, we introduce a quadrupolar Heisenberg term to the Hamiltonian:
\begin{align}
\mathcal{H}_{\mathcal{Q}} = J_{\mathcal{Q}} \sum_{\langle i j \rangle} \vec{\tau}_{f,i} \cdot \vec{\tau}_{f,j}
\end{align}
where $\vec{\tau}_{f,i} = \frac{1}{2} f^\dagger_{i\alpha}\vec{\tau}_{\alpha \beta} f_{j\beta}$ is the $\Gamma_3$ pseudospin and $\vec{\tau}$ a vector of Pauli matrices.  Within the present mean-field theory, the decouplings into the two different quadrupolar moments $Q_\mu \propto 3z^2 - r^2$ and $Q_\nu \propto x^2 - y^2$ are degenerate at $\vec{B}=0$.  In finite field $B \parallel z$, $Q_\mu$ is favored, and so we use it here to examine the state which competes most strongly with hastatic order. We therefore choose the specific mean-field decoupling of this interaction to yield an AFQ order parameter $\mathcal{Q}$ along the $z$ axis, with the resultant mean-field Hamiltonian
\begin{align}
\mathcal{H}_{\mathcal{Q},MF} = &-\mathcal{Q} \sum_{k} \Big[   (f^\dagger_{k+Q,+} f_{k+}-f^\dagger_{k+Q,-} f_{k-}) + H. c.   \Big]  \notag \\
 &+ \frac{ 3\mathcal{N} \mathcal{Q}^2}{J_Q}.
\end{align}
Experiments on PrIr$_2$Zn$_{20}$ have detected $Q_\nu$ AFQ order, but the qualitative features of our calculated phase diagrams are expected to remain the same with this choice as well.  Adding this $\mathcal{H}_{\mathcal{Q},MF}$ to our FH mean-field Hamiltonian, we obtain the free energy and solve the saddle point equations, $\partial \mathcal{F}/\partial \lambda = \partial \mathcal{F}/\partial \mu = \partial \mathcal{F}/\partial b = \partial \mathcal{F}/\partial \mathcal{Q} =0$; note that the AFQ mean-field is not justified within the $SU(N)$ large-$N$ limit, and this mean-field theory is therefore not controlled.

\begin{figure}[ht]
\vspace*{-0cm}
 \includegraphics[scale=0.45]{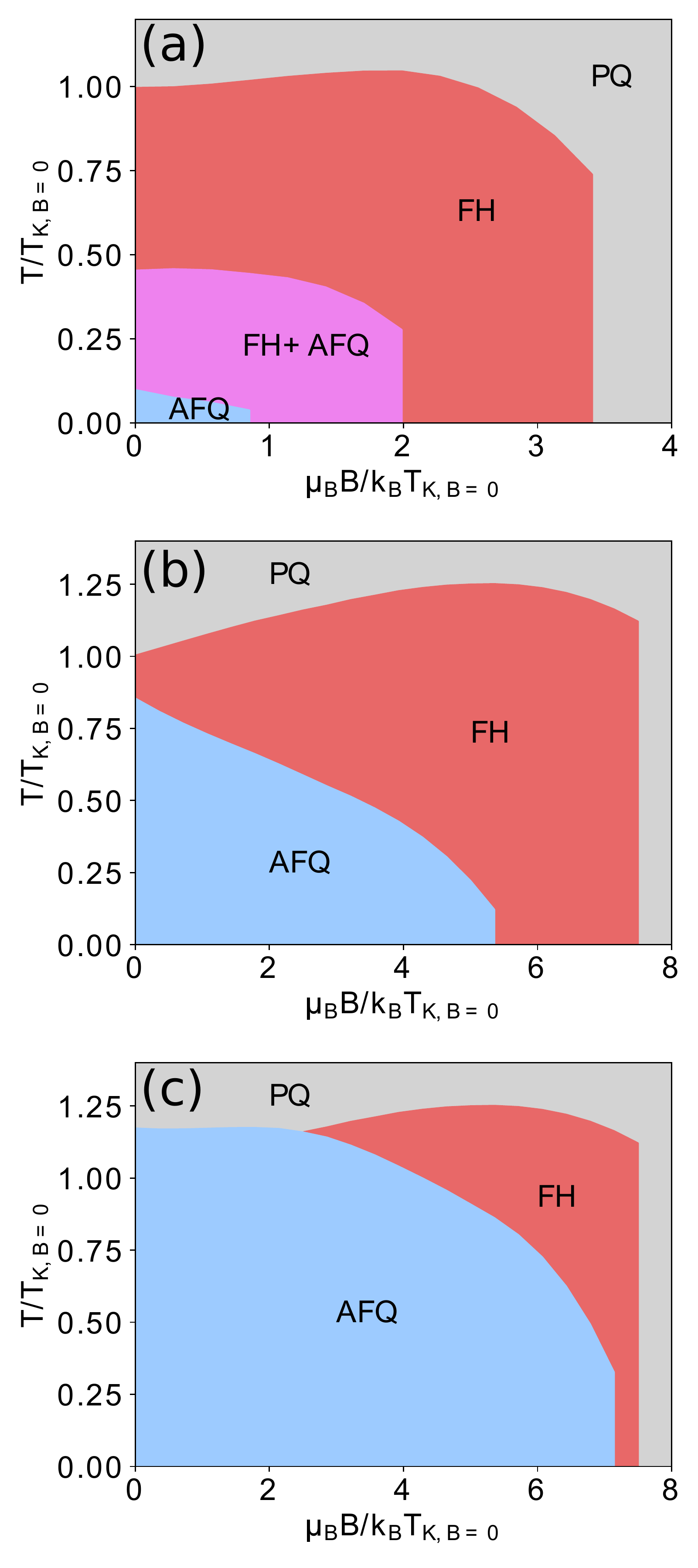}%
\vskip -0.2cm
 \caption{Three example mean-field phase diagrams for FH (red) and AFQ (blue) orders in magnetic field, for three different strengths of quadrupolar exchange coupling: (a) $J_Q\!=\!0.38t$, (b) $J_Q\!=\!1.6t$, (c) $J_Q\!=\!2t$.  The transition to FH order, $T_K$ initially rises with magnetic field and is then suppressed, ultimately becoming a first-order transition in field. Other parameters for (a): $t\!=\!1$, $V\!=\!0.8$, $\Delta E\!=\!5.5$, $\Delta\!=\!4.8$, $n_{c,0}\!=\!1.9$; other parameters for (b),(c): $t\!=\!1$, $V\!=\!0.9$, $\Delta E\!=\!5.5$, $\Delta\!=\!20$, $n_{c,0}\!=\!1.6$. \label{fig:PD}}
\end{figure}

At zero and small fields, the Pr(Ir,Rh)$_2$Zn$_{20}$ compounds order quadrupolarly, giving way to heavy Fermi liquid behavior at intermediate fields \cite{Onimaru2016b,Yoshida2017}.  We qualitatively reproduce this behavior in our self-consistently calculated mean-field phase diagrams for FH and AFQ ansatze in magnetic field, shown in Fig.~\ref{fig:PD} for three different sets of $J_Q$ and $n_c^0$ parameters that capture three possible behaviors.  Near half-filling, for small $J_Q$, there is a large coexistence region with FH order at low fields [Fig.~\ref{fig:PD}(a)].  By increasing $J_Q$ and moving away from half-filling, the coexistence region can be made to disappear, replaced by a direct transition between the FH and AFQ phases [Fig. \ref{fig:PD}(b,c)], which is reminiscent of the experimental result, with FH order explaining the heavy Fermi liquid region in intermediate fields. 

Due to the linear coupling of the magnetic field to the FH moments, we expect magnetic field to initially favor FH order, leading to increased hybridization and higher $T_K$.  This increase is seen in all three subplots of Fig. \ref{fig:PD}, most noticeably in part (b).  However, as stronger fields split the $\Gamma_3$ doublet, the Kondo screening and thus FH order are eventually destroyed, as seen in the mean-field calculation.  AFQ order is also suppressed as a function of $B$, leading to a first-order transition between it and FH order at a critical field $B_c$.  Note that while the mean-field theory always finds a first order transition between FH and PQ states at low temperatures and high fields, we do not expect this to necessarily hold beyond the mean-field level.

\section{Landau theory \label{sec:landau}}

Large-$N$ theories have been extremely useful in understanding Kondo physics at low temperatures, where they work well \cite{hewson}.  However, these theories are known to have issues near the Kondo temperature -- most notably in predicting a phase transition in the single channel Kondo effect.  As the single channel ``order parameter'' $\langle b\rangle$ breaks only the emergent gauge symmetry, Elitzur's theorem prevents it from ordering; indeed, $1/N$ corrections show that the bosonic expectation value is not long range ordered and wash out the phase transition \cite{Read1985}.  In the two channel case, our bosonic order parameter $\langle b_\mu \rangle$ breaks a real symmetry in addition to the gauge symmetry and so the transition must survive.

The key question here is exactly how the hastatic order parameter breaks the symmetry.  Is the order parameter really spinorial, that is, described by a spinor (double group) representation?  Or is it vectorial, like most known order parameters?  We know the answer in the large-$N$ limit, where our mean-field theory is strictly correct. The infinite-$N$ order parameter is spinorial, in the $\Gamma_7$ double group irreducible representation. In this limit, the order parameter does not couple linearly to the magnetic field, which has a $\Gamma_4$ symmetry, and there is always a phase transition into the FH phase at $T_K$, even in finite field.  Note that the usual Kondo ``order parameter'' also survives in the large-$N$ limit, but is washed out with $1/N$ corrections. There are strong reasons to believe that the $\Gamma_7$ nature of the order parameter does not survive the $1/N$ corrections that wash out $\langle b\rangle$ in the single channel case.  The first is that the conjugate field to the FH order parameter along $\hat z$ is the breaking of the channel symmetry $\delta J = \frac{J_\uparrow - J_\downarrow}{2}$, where $J_\sigma$ is the Kondo coupling in each conduction electron channel.  As soon as the channel symmetry is broken, a heavy Fermi liquid develops in the strongest channel below a crossover scale $T_K$; this heavy Fermi liquid \emph{is} FH order.  We can see that $\delta J$ is the conjugate field to the composite order parameter by rewriting the two-channel Kondo coupling in terms of $\Psi_z = \langle \tilde{\sigma}c\dg_{\sigma}\vec{\tau}c_\sigma \cdot \vec{\tau}_f \rangle$ \cite{Komijani2018}:
\begin{equation}
H_K = J \sum_\sigma c\dg_{\sigma}\vec{\tau}c_\sigma \cdot \vec{\tau}_f + \delta J \sum_\sigma \tilde{\sigma} c\dg_{\sigma}\vec{\tau}c_\sigma \cdot \vec{\tau}_f.
\end{equation}
Of course, we can also write down conjugate fields that couple to composite orders in the $x$ and $y$ directions, for $\vec{\Psi} = \langle c\dg \vec{\sigma} [\vec{\tau} \cdot \vec{\tau}_f] c\rangle$ and in fact, overall the composite order parameter forms an $SO(5)$ order parameter that includes composite pairing, $\Phi = \langle c\dg i\sigma_2 [(i\tau_2 \vec{\tau})\cdot \vec{\tau}_f]c\dg\rangle$ and $\Phi\dg$ order parameters \cite{Hoshino2014,Flint2008}, although this $SO(5)$ symmetry is broken in the cubic Anderson model. In cubic symmetry, $\vec{\Psi}$ belongs to the $\Gamma_{4}$ representation, and so the conjugate field $\vec{\delta J}$ will also have $\Gamma_{4}$ symmetry.  The Landau order parameter at the phase transition must therefore be the $\Gamma_{4}$ composite order parameter, $\vec{\Psi}$ which behaves like $\langle b\dg \vec{\sigma} b\rangle$, and not the $\Gamma_7$ spinorial order parameter $\langle b_\mu \rangle$.   We can also understand the vector nature of the order parameter by appealing to the Higgs mechanism in the Kondo effect that locks together the internal and external gauge fields and gives charge to the composite fermions.  For the single-channel Kondo effect, the phase of the hybridization plays the role of the Goldstone boson, and is absorbed to make the difference between internal and external gauge fields heavy.  For the two-channel Kondo effect, we have an $SU(2)$ spinor,
\begin{equation}
    b_j = |b_j|\mathrm{e}^{i\chi_j}\left( \begin{array}{c} \cos \theta_j \mathrm{e}^{i\phi_j/2} \\ \sin \theta_j \mathrm{e}^{-i\phi_j/2}\end{array}\right).
\end{equation}
Here, the Higgs mechanism absorbs the overall phase $\chi_j$, leaving an $SO(3)$ order parameter defined by two angles (and the overall amplitude)\cite{piers}. 

The correct FH order parameter is then $\vec{\Psi}$, which has $\Gamma_4$ symmetry and couples linearly to the magnetic field, like a ferromagnet.  There are several key differences between FH and ferromagnetic orders though, which it is important to keep in mind.
\begin{itemize}
    \item The composite order parameter, $\vec{\Psi}$ and the mixed valent moment, $\langle b\dg \vec{\sigma}b\rangle$ have the same symmetry, but will typically have very different magnitudes.  In the Kondo limit of integer valence, $\langle b\rangle = 0$, and yet $|\Psi|$ will still be large.  Therefore, the coupling of $\vec{\Psi}$ to external field, $\vec{h}$ will typically be quite small.  In this sense, while very close to the phase transition FH order will look like a ferromagnet, with diverging susceptibility, further from the phase transition, it will look like the spinorial order parameter, with linear magnetization,  instead of the square-root behavior of a ferromagnet, for example.
    \item The composite order parameter does not commute with the Hamiltonian, although it has been shown to retain the quadratic Goldstone modes \cite{Hoshino2015,piers}.
    \item The development of a hybridization gap is associated with hastatic order, with the gap magnitude growing as $\sqrt{|\Psi|}$.  Additionally, the originally neutral pseudofermions $f_\alpha$ pick up electric charge via a Higgs mechanism and become part of the Fermi surface, and so a discrete change in Fermi surface volume is expected across the FH transition.  
\end{itemize}

If we want to understand the behavior of FH order near the phase transition, we need to examine the Landau theory of the composite order parameter $\vec{\Psi}$, keeping in mind the weak linear coupling to external field. The Landau theory should really be thought of as capturing how $1/N$ corrections will modify the behavior near the transition, while the low temperature physics is still expected to be well-described by our mean-field theory.

To explore these consequences in detail, and the effect on the thermodynamic responses, we consider a simple Landau theory.  As AFQ order is a natural competitor for FH order, we will compare the behavior of FH and AFQ orders. As the theory is complex, we introduce it in stages, but our goal is a full theory of the interplay of FH and AFQ orders in both magnetic field and strain. Here, we neglect the possible octupolar order of the $\Gamma_3$ doublet; a Landau theory of quadrupolar and octupolar orders, and their interaction with external field and strain was recently developed \cite{Lee2018,Patri2018}.


\subsection{Ferrohastatic order}

The allowed terms in a Landau theory are found by considering products of the representations of the various order parameters, and taking all the invariant ($\Gamma_1$) terms of each order.   The local site symmetry of the Pr atoms is known to be $T_d$ \cite{Niemann1995}, and so the appropriate group for the composite order parameter, $\vec{\Psi}$ is $T_d \times \tau$, where $\tau$ is time-reversal symmetry.  $\vec{\Psi}$ is described by the same $\Gamma_{4u}$ irreducible representation as the external magnetic field $\vec{h}$, and so magnetic field is expected to smear out the phase transition into a crossover.  Here we use $u/g$ to indicate odd/even behavior under time-reversal.  We first find all quadratic terms in both $\vec{\Psi}$ and $\vec{h}$:
\begin{align}
\Gamma_{4u} \otimes \Gamma_{4u} & = \Gamma_{1g} \oplus \Gamma_{3g} \oplus \Gamma_{4g} \oplus \Gamma_{5g}\cr
& = |\Psi|^2 \oplus \overrightarrow{\Psi^2}_{\Gamma_3} \oplus \cdot \oplus \overrightarrow{\Psi^2}_{\Gamma_5}\cr
& \qquad \mathrm{or}\cr
& = |h|^2 \oplus \overrightarrow{h^2}_{\Gamma_3} \oplus \cdot \oplus \overrightarrow{h^2}_{\Gamma_5}\cr
& \qquad \mathrm{or}\cr
& = \vec{h} \cdot \vec{\Psi} \oplus \overrightarrow{h \Psi}_{\Gamma_3} \oplus \vec{h} \times \vec{\Psi} \oplus \overrightarrow{h \Psi}_{\Gamma_5}
\end{align}
Here $\overrightarrow{\phi^2}_{\Gamma_3}= (\frac{1}{\sqrt{3}}\left[3\phi_z^2-|\phi|^2\right],\phi_x^2-\phi_y^2)$ and $\overrightarrow{\phi^2}_{\Gamma_5}=(\phi_x \phi_y,\phi_y \phi_z,\phi_z \phi_x)$, for $\phi = h$ or $\Psi$ and the $\Gamma_{4g}$ $\vec{\phi}\times\vec{\phi}$ term vanishes.  We also have the mixed terms $\overrightarrow{h \Psi}_{\Gamma_3} = (\frac{1}{\sqrt{3}}\left[3h_z \Psi_z-|h||\Psi|\right],h_x\Psi_x-h_y \Psi_y)$ and $ \overrightarrow{h \Psi}_{\Gamma_5} =(h_x \Psi_y + h_y \Psi_x,h_y \Psi_z + h_z \Psi_y,h_z \Psi_x +h_x \Psi_y)$.  These terms can be used to construct the allowed fourth order terms with $\Gamma_1$ symmetry; there are no allowed third order terms due to the time-reversal symmetry breaking nature of the order parameter.  And so we construct the Landau theory,
\begin{equation}
F_\Psi = \alpha_\Psi |\Psi|^2 + \frac{u_\Psi}{2} |\Psi|^4 - \lambda \vec{h}\cdot \vec{\Psi} + u_{h\Psi}|h|^2|\Psi|^2
\end{equation}
Here, we neglect several fourth order terms: there are terms that pin the hastatic spinor, $\overrightarrow{\Psi^2}_{\Gamma_3} \cdot \overrightarrow{\Psi^2}_{\Gamma_3}$ and $\overrightarrow{\Psi^2}_{\Gamma_5} \cdot \overrightarrow{\Psi^2}_{\Gamma_5}$ -- either to the $[111]$ or $[100]$ directions, respectively.  Microscopic calculations show these pinning terms to be quite weak, and the $\lambda$ magnetic field coupling will quickly overwhelm them to pin the hastatic order parameter along the external field direction. We also drop several second order terms in both $h$ and $\Psi$ for the same reason.  Remember that the composite order parameter $\vec{\Psi}$ is nonzero even in the Kondo limit where the mixed valent moment $b\dg \vec{\sigma} b$ vanishes, and so the coupling to external field will be extremely small when the materials are near integral valence, as is likely the case for the Pr-based materials. 

From this Landau theory, we can already see that FH order is a type of ferromagnetism, but with a peculiarly weak linear coupling to external field.  We expect a divergence in the magnetic susceptibility at $T_K$ in zero field, although the coefficient [$\lambda^2/(2 \alpha_\Psi)$] is small.  For finite fields, the susceptibility is nearly constant above the transition and then develops a linear component, $d\chi/dT \approx 2 u_{h \Psi} \alpha_\Psi/u_\Psi$ below the transition, where this equation is  exactly true if $\lambda  = 0$.  For $\lambda = 0$, the magnetic moment grows linearly in temperature below $T_K$, while finite $\lambda$ leads to the typical square root behavior in zero field.  For small $\lambda$ and finite $h$, the linear behavior is still evident slightly away from the transition and the field smears out the kink.  All signatures of the phase transition, like the specific heat jump, will be similarly smeared out, governed by $\lambda$.

\subsection{Ferrohastatic order and coupling to strain}

As the $\Gamma_3$ doublet has quadrupolar components $O_2^0$ and $O_2^2$, including coupling to strain is extremely important.  There are five strain components, 
\begin{align}
\overrightarrow{\epsilon}_{\Gamma_{3g}}& = (\epsilon_\mu, \epsilon_\nu) = \left(\frac{1}{\sqrt{3}}[2\epsilon_{zz} - \epsilon_{xx}-\epsilon_{yy}], \epsilon_{xx} - \epsilon_{yy}\right)\cr
\overrightarrow{\epsilon}_{\Gamma_{5g}} & = (\epsilon_{xy},\epsilon_{yz},\epsilon_{zx}),
\end{align}
The first two components, $\overrightarrow{\epsilon}_{\Gamma_{3g}}$ couple linearly to the possible ferroquadrupolar (FQ) orderings of the $\Gamma_3$ doublet:  $\vec{R} = (R_\mu, R_\nu)$.  Here $R_\mu = \langle O_2^0 \rangle$ and $R_\nu = \langle O_2^2\rangle$ are the two possible FQ orders.  The $\Gamma_5$ components will couple to the $\Gamma_5$ combinations of $\Psi$ and $h$.  However, $\overrightarrow{h\Psi}_{\Gamma_5}$ requires $\vec{h} \perp \vec{\Psi}$, which is forbidden by the pinning of the hastatic spinor to the external field direction, and so we neglect these strain components entirely. 

The elastic free energy for $\overrightarrow{\epsilon}_{\Gamma_{3g}}$ is,
\begin{align}
F_{el} = \frac{c_{11}-c_{12}}{2} (\epsilon_\mu^2+\epsilon_\nu^2) - g_3 \vec{R} \cdot \overrightarrow{\epsilon}_{\Gamma_{3g}},
\end{align}
where $c_{11}$ and $c_{12}$ are elastic coefficients, and we take $g_3 < 0$, as in PrIr$_2$Zn$_{20}$ \cite{Worl2018}.
We can then integrate out the strain,
\begin{equation}
\epsilon_{\mu,\nu} = \frac{g_3}{c_{11}-c_{12}} R_{\mu,\nu},
\end{equation}
and work directly with the ferroquadrupolar order parameters.  We can again use group theory to determine the symmetries of different combinations of $\vec{R}$:
\begin{align}
\Gamma_{3g} \otimes \Gamma_{3g} & = \Gamma_{1g} \oplus \Gamma_{2u} \oplus \Gamma_{3g} \cr
& = |R|^2 \oplus \cdot \oplus \overrightarrow{R^2}_{\Gamma_3} \cr
&= |R|^2 \oplus \cdot \oplus (R_\nu^2-R_\mu^2,2 R_\mu R_\nu).
\end{align}
Note that there is no $\Gamma_{2u}$ term -- the original $\Gamma_3$ multiplets have $\Gamma_{3g}$ quadrupolar orders and $\Gamma_{2u}$ octupolar order, but this octupolar order cannot be constructed from the time-reversal invariant quantities here and must be treated independently, as has been done recently \cite{Patri2018}.

The Landau theory for FQ order is,
\begin{equation}
F_R = \alpha_R |R|^2 + \frac{u_R}{2} |R|^4 + v_R \vec{R} \cdot \overrightarrow{R^2}_{\Gamma_3} - \gamma_R \vec{R}\cdot \overrightarrow{h^2}_{\Gamma_3}
\end{equation}
Here, we assume $\alpha_R > 0$ to forbid intrinsic FQ order; it will be induced by both FH and AFQ orders, as well as finite field. The third order clock term pins the FQ order parameter to the lattice. We know from single-ion physics that magnetic field favors FQ order via induced magnetic moments, and so $\gamma_R > 0$.

Finally, we can couple the two orders:
\begin{equation}
F_{\Psi R} = \kappa_R \vec{R} \cdot \overrightarrow{h \Psi}_{\Gamma_3}+ \nu_R \vec{R} \cdot \overrightarrow{\Psi^2}_{\Gamma_3} + u_{\Psi R}|\Psi|^2|R|^2
\end{equation}
Note that FQ order appears to develop immediately with FH order, due to $\nu_{R}$, however, $\nu_{R}$ vanishes in our microscopic theory due to the nodal structure of the hybridization. $\kappa_R$ induces FQ order whenever both $\Psi$ and $h$ are nonzero, and this term is nonzero in the microscopic theory, although likely to be small, just as $\lambda$ is.

The thermal expansion $\alpha$ and magnetostriction $\lambda$ are defined in terms of the fractional change in length, $\Delta L/L$ along some direction, which is in turn proportional to strain,
\begin{align}
\frac{\Delta L}{L} \biggr \rvert_{z} & = \frac{1}{3} \epsilon_B + \frac{1}{\sqrt{3}} \epsilon_\mu\cr
\frac{\Delta L}{L} \biggr \rvert_{x} & = \frac{1}{3} \epsilon_B - \frac{1}{2\sqrt{3}} \epsilon_\mu + \frac{1}{2} \epsilon_\nu,
\end{align}
where $\epsilon_B$ is the symmetric volume strain.  For specificity, we consider the magnetic field to be along $\hat z$, and define,
\begin{align}
\alpha_{\parallel} & = \frac{1}{L}\frac{d\Delta L}{dT} \biggr \rvert_{z}; \quad \alpha_{\perp} = \frac{1}{L}\frac{d\Delta L}{dT} \biggr \rvert_{x} \cr
\lambda_{\parallel} & = \frac{1}{L}\frac{d\Delta L}{dh} \biggr \rvert_{z}; \quad \lambda_{\perp} = \frac{1}{L}\frac{d\Delta L}{dh} \biggr \rvert_{x}.
\end{align}

None of the order parameters couple linearly to the bulk $\epsilon_B$, and as we consider the  $R_\mu = O_2^0$ FQ order (so $\epsilon_\nu = 0$), motivated by experiments on PrIr$_2$Zn$_{20}$ \cite{Worl2018}, the relationships $\alpha_\parallel + 2 \alpha_\perp = 0$ and $\lambda_\parallel + 2 \lambda_\perp = 0$ always hold. Note that these are likely violated beyond Landau theory, where changes in $f$-electron valence typically result in volume changes.  Here, we focus on the parallel components, with the perpendicular components understood to be given by these relations.  

\begin{figure}
 \includegraphics[scale=0.55]{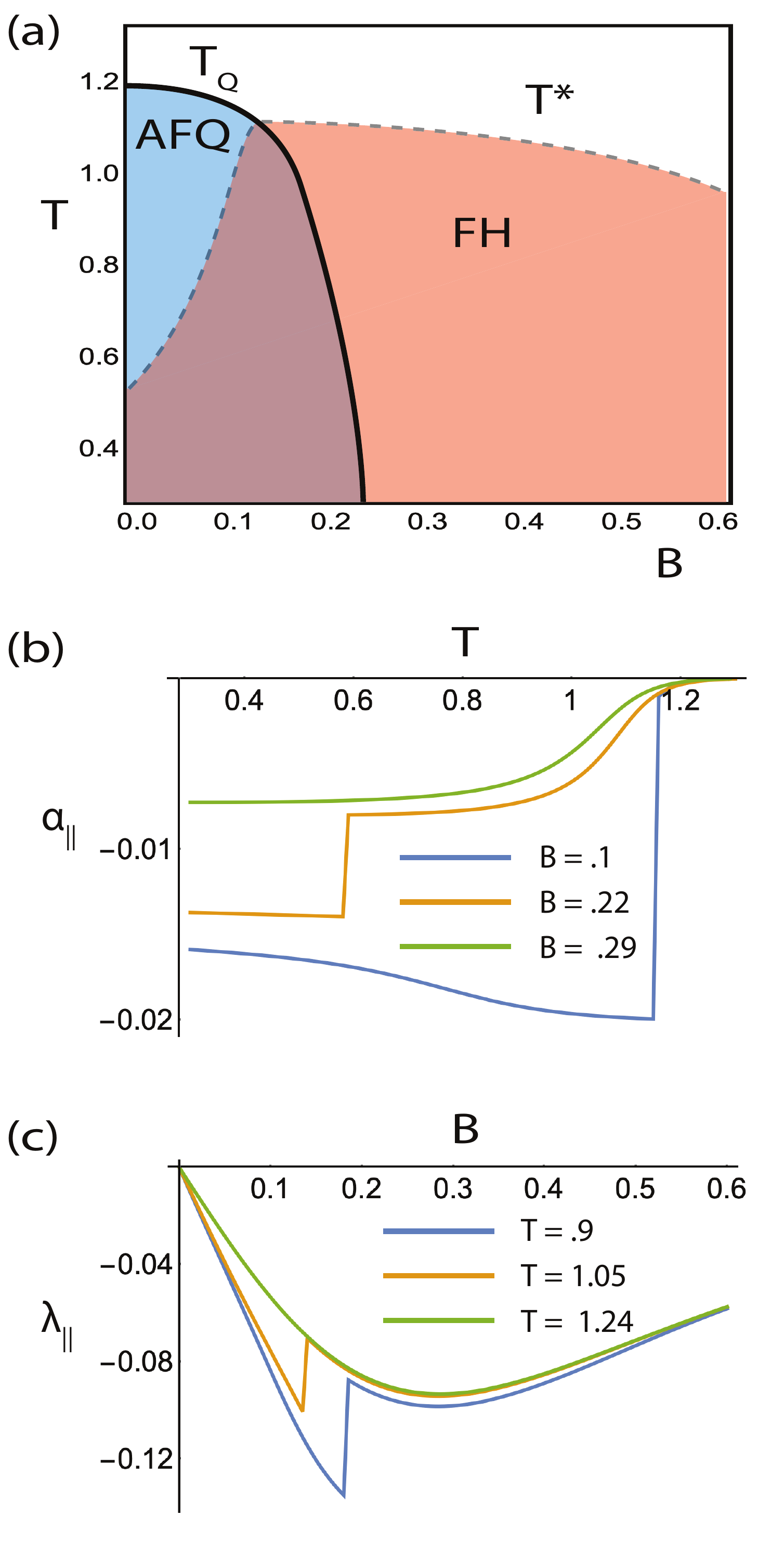}%
 \caption{Signatures of ferrohastatic and antiferroquadrupolar orders. (a) Example phase diagram in temperature (T) and magnetic field along $\hat z$ (B). $T_Q$ indicates the phase transition into AFQ order, while $T^*$ is the crossover scale for FH order; $T^*$ becomes a phase transition only for $B = 0$. (b) Thermal expansion ($\alpha_\parallel$) as a function of temperature for three different fields. There is a sharp negative jump upon entering the AFQ phase, and a smeared out jump upon cooling through $T^*$ which is negative for $T^*(B) > T_Q(B)$ and positive otherwise. Note that the Landau theory captures $\alpha_\parallel$ only near the transition - far from the transition, the microscopic physics will lead to non-monotonic behavior and eventually $\alpha_\parallel$ goes to zero as $T \rarrow 0$.  (c) Magnetostriction ($\lambda_\parallel$) as a function of magnetic field for three different temperatures.  There is a negative jump upon exiting the AFQ order that grows with decreasing temperature, but the signature at $B^*$ is completely smeared out. Parameters are given in the text.
 \label{fig:Landau}}
 \end{figure}

The coupling of FH and FQ orders leads to negative jumps in both the thermal expansion and magnetostriction at the transition into hastatic order if either $h = 0$ or the couplings $\lambda$ and $\kappa_R$ are zero.  If $\lambda$ and $\kappa_R$ are nonzero and $h$ is finite, as is expected experimentally, the jumps are slightly smeared. We show some examples in Fig.~\ref{fig:Landau}, where we consider both AFQ and FH orders.

\subsection{Comparison with antiferroquadrupolar order}

A similar analysis for the $\Gamma_3$ AFQ order in magnetic field yields \cite{Lee2018},
\begin{align}
F_Q & = \alpha_Q (T-T_Q)|Q|^2 + \frac{u_Q}{2} |Q|^4 + w_Q |Q|^6\cr
& +v_Q (-Q_\nu^3+3 Q_\mu^2 Q_\nu)^2+u_{Qh} |Q|^2|h|^2\cr
& + v_{Qh} \overrightarrow{Q^2}_{\Gamma_3} \cdot \overrightarrow{h^2}_{\Gamma_3} 
\label{eq:landauFS1}
\end{align}
where $Q_\mu = \langle O^0_2 \rangle_A - \langle O^0_2\rangle_B$ and $Q_\nu = \langle O^2_2 \rangle_A - \langle O^2_2 \rangle_B $ are the AFQ order parameters comprising the $\Gamma_3$ doublet, and we keep only the lowest order symmetry-breaking terms.  $A$, $B$ are the two sublattices of the diamond structure, which describes the arrangement of Pr ions.  The sixth order term, $v_Q$ is the square of the third order in $Q$ term with $\Gamma_2$ symmetry. 
The coupling to FQ order is given by,
\begin{align}
F_{QR} & = \rho \vec{R} \cdot \overrightarrow{Q^2}_{\Gamma_3} +u_{QR} |R|^2|Q|^2\label{eq:landauFS2}
\end{align}
$\overrightarrow{Q^2}_{\Gamma_3}$ is defined identically to $\overrightarrow{R^2}_{\Gamma_3}$, and couples linearly to the FQ order parameter $\vec{R}$.  We neglect higher order terms as subdominant.

Here, $\rho$ leads to jumps in thermal expansion and magnetostriction that can take either sign, although $\rho < 0$ is indicated by the experimental results, wherein $R_\mu$ is induced by $Q_\nu$\cite{Iwasa2017,Worl2018}. 

\subsection{Coupling ferrohastatic and antiferroquadrupolar orders}

For completeness, the interactions between FH and AFQ order are captured in,
\begin{equation}
F_{\Psi Q} = u_{\Psi Q}|Q|^2|\Psi|^2 + \kappa_Q \overrightarrow{h \Psi}_{\Gamma_3} \cdot \overrightarrow{Q^2}_{\Gamma_3} + \nu_Q \overrightarrow{\Psi^2}_{\Gamma_3} \cdot \overrightarrow{Q^2}_{\Gamma_3}
\end{equation}
Note that $\nu_Q$ also vanishes in our microscopic theory.  AFQ and FH orders suppress one another, and can either coexist (for sufficiently small $u_{\Psi Q}$) or phase separate via a first order transition (for larger $u_{\Psi Q}$).  We find both cases in our microscopic theory above, for different values of the AFQ coupling, and show a Landau theory example in the next section.

\subsection{Example phase diagram and thermodynamics}

In Fig. \ref{fig:Landau} (a), we show one possible phase diagram in temperature and field. Here, we choose our Landau parameters to roughly reproduce the experimental phase diagram.  The AFQ order parameter is $Q_\nu$, while the FQ order parameter (not shown) is $R_\mu$, and the FH order parameter, $\vec{\Psi}$ points along $\hat z$.  We similarly choose the magnetic field $\vec{h} = B \hat z$.  The Landau parameters are:
\begin{align}
T_K & = 1.1, \alpha_\Psi = 1, u_\Psi = 7, \lambda = 0.05, u_{h\Psi} = 1\cr
\alpha_R & = 1, u_R = 3, v_R = 0, \gamma_R = 1 \cr
\kappa_R & = -0.05, \nu_R = -0.25, u_{\Psi R}=0.5 \cr
T_Q & = 1.2, \alpha_Q = 1, u_Q = 3.5, v_Q = 0, u_{Qh} = 4\cr
\rho & = -0.5, u_{QR} = 0, u_{\Psi Q} =6, \kappa_Q = 0, \nu_Q = 0
\end{align}
Note that  $\lambda$ and $\kappa_R$ are nonzero but small, to reflect the smallness of the moment relative to the magnitude of the composite order parameter, $\vec{\Psi}$ for nearly integral valence.  For any finite $B$, the FH phase transition is smeared out by these parameters.  FQ order only turns on via interactions with other order parameters and magnetic field.  The parameters here were chosen to roughly reproduce the single ion behavior of the magnetostriction in magnetic field.  The signs of $\rho$ and $\nu_R$ are chosen to reproduce the negative jump in the thermal expansion seen in PrIr$_2$Zn$_{20}$ \cite{Worl2018}.  $\nu_R$ is zero in our microscopic theory, but is generically allowed to be nonzero by symmetry; we take it to be small, but negative to match the sign of the  experimental thermal expansion jump.  FH and AFQ orders have similar zero field transition temperatures (which requires fine-tuning, of course), and they strongly repel one another via $u_{\Psi Q}$.  We otherwise set $\kappa_Q$ and $\nu_Q$ to zero for simplicity.

\section{Signatures of ferrohastatic order \label{sec:signatures}}

Fundamentally, FH order is a heavy Fermi liquid with a spinorial hybridization that breaks the channel symmetry.  As such, it has two types of signatures: heavy Fermi liquid behavior, where half of the conduction electrons hybridize with the local moments and half remain unhybridized, and symmetry breaking signatures, including magnetic moments and thermodynamic signatures.

\subsection{Heavy Fermi liquid behavior}

\begin{figure}[ht]
\vspace*{-0cm}
 \includegraphics[scale=0.34]{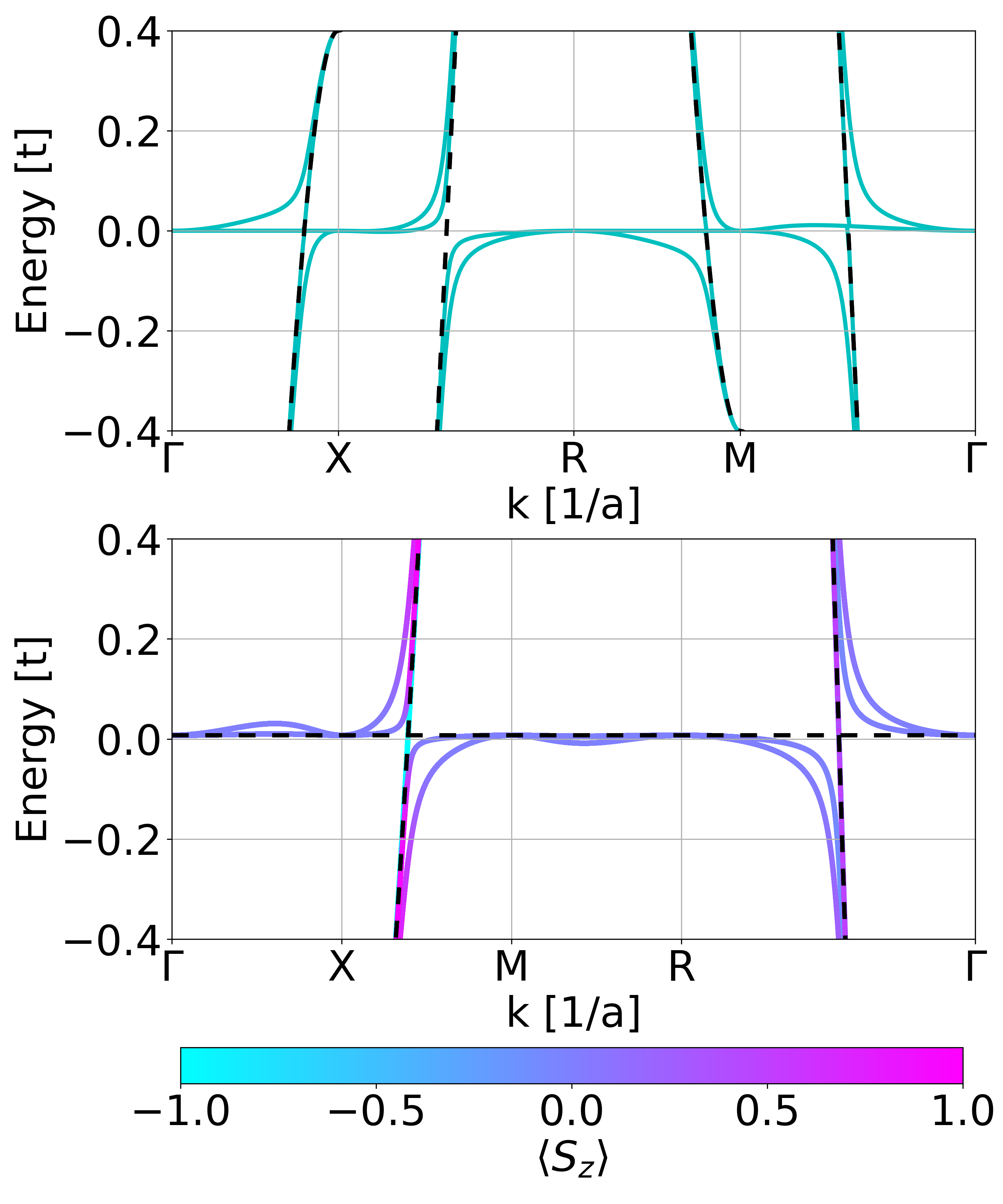}%
\vskip -0.2cm
 \caption{\small Two example FH dispersions to illustrate the nature of the hybridization. (a) Example dispersion of heavy quasiparticles in which all conduction bands are hybridized, as seen from the $X--R$ cut, which is not a high symmetry line.  (b) Example dispersion in which one conduction electron band always remains unhybridized; here, we also use the color scale to show the spin polarization of the heavy bands.  The unhybridized $c$- and $f$-bands (dashed black lines) and hybridized bands (thick solid lines) are plotted along high symmetry lines in the cubic Brillouin zone, near $E_F=0$. The color indicates the projection of spin along the z-axis; note that $\langle S_z\rangle = 0$ merely implies that the spins lie in the $xy$-plane. Parameters for (a): $t\!=\!1$, $\mu\!=\!0$, $\eta_c\!=\!0.2$, $\lambda\!=\!0$, $V\!=\!\eta_v\!=\!0.1$, $b_\mu\!=\!(1,0)$; (b): $t\!\!=\!\!\eta_c\!\!=\!\!V\!\!=\!\!1$, $\eta_v \!\!=\!\! -\sfrac{1}{3}$, $\Delta E\!\!=\!\! 5.5$, $n_{c,0}\!\!=\!\!1.6$ [$\lambda$, $b_\mu$, $\mu$ determined self-consistently for (b)]. \label{fig:dispersions}}
\end{figure}

In the simplest cases, FH order is a half-heavy Fermi liquid -- one band of conduction electrons hybridizes and becomes heavy, while the other remains light. Along high symmetry lines, or for the simple case of $\eta_c = 1, \eta_V = 1$, this is true: one light band remains completely unmodified, while the other hybridizes and becomes heavy.  In more generic cases, however, both bands become hybridized due to the strong spin-orbit coupling, although one is more strongly modified [Fig.~\ref{fig:dispersions}(a)].  In the simple two-channel Kondo model, for example, the spin-up conduction electrons hybridize and form a heavy band, with a hybridization gap, while the spin-down conduction electrons remain unhybridized and ungapped to form a light band\cite{Zhang2018}.  In the more realistic model considered here, the spin-orbit coupled hybridization means that the spin structure of the heavy band varies throughout the Brillouin zone, as shown in Fig. \ref{fig:dispersions}(b).

The heavy band dominates the thermodynamic properties, and FH order has all the traditional signatures of heavy Fermi liquids, including a large Sommerfeld coefficient and $A T^2$ resistivity. The two bands will have very different effective masses, which can be probed by quantum oscillations. The resulting ``half''-hybridization gap should be observed in angle-resolved photoemission spectroscopy (ARPES), scanning-tunneling microscopy (STM), and optical conductivity measurements. The optical conductivity sum rule, $n(\omega)=\frac{m}{e^2}\int_0^\infty \frac{d\omega'}{\pi} \sigma_1 (\omega')$, will have a kink at approximately half of the total weight, as a direct consequence of the half-hybridization gap pushing spectral weight of the initial Drude peak above the direct gap.  As the $\Gamma_8$ form-factors mix spin and orbital angular momentum, the physical spin structure of both hybridized and unhybridized bands varies throughout momentum space, as shown in Fig. \ref{fig:dispersions}(b); this structure could be detected in spin-resolved ARPES.

\begin{figure}[ht]
\vspace*{-0cm}
 \includegraphics[scale=0.5]{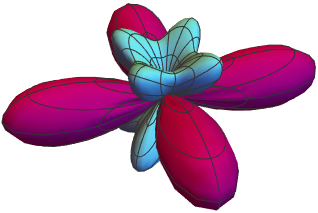}%
\vskip -0.2cm
 \caption{\small  The symmetry breaking hybridization gap has a  $\Gamma_3 +$ angular dependence for $\hat b \parallel \hat z$.  Blue and red indicate positive and negative values, respectively.\label{fig:symmbreakgap}}
\end{figure}

In zero magnetic field, there are generically two types of hybridization gaps: symmetric gaps that only depend on the amplitude of the hastatic spinor, $\Tr [\hat{b}^\dagger \hat{b}]$ and symmetry-breaking gaps that depend on the direction, $\Tr [\hat{b}^\dagger \vec{\mu} \hat{b}]$, where $\vec{\mu}$ is a vector of channel Pauli matrices. Spin-orbit coupling implies that the symmetry breaking gaps break both $SU(2)$ and cubic symmetries, which we show in Fig. \ref{fig:symmbreakgap}(b) for special parameters that allow an analytic form of the dispersion. Here, $\hat b || \hat z$ and the gap has $\Gamma_3 +$ symmetry (see Appendix A for details).  Note that in mean-field theory, both gaps develop via a phase transition at $T_K$, but fluctuations will allow the non-symmetry-breaking gap to develop as a crossover at a higher $T^*$, along with the heavy Fermi liquid signatures.

\subsection{Symmetry breaking signatures}

\begin{figure}[ht]
 \includegraphics[scale=0.5]{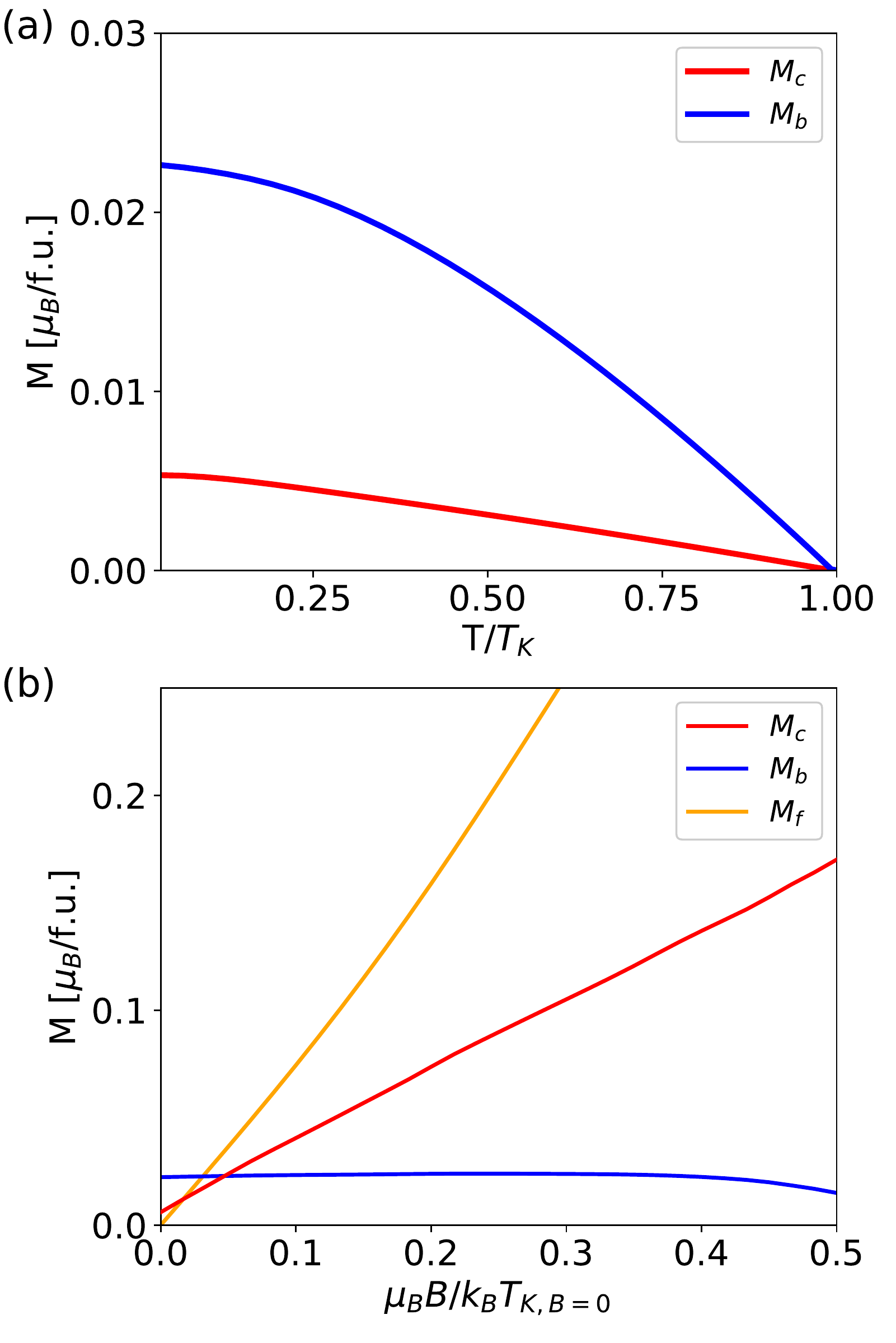}%
 \caption{(a) Behavior of conduction electron (red) and $\Gamma_7$ (blue) moments with temperature; note that each turns on linearly in field, in contrast to ferromagnetic order. (b) Conduction electron (red), 4f$^3$ $\Gamma_7$(blue) and 4f$^2$ $\Gamma_3$(orange)
moments as a function of field; in FH order, all moments are strictly parallel to the applied field.  Parameters are $t=1$, $\eta_c=1$, $V=0.8$, $\eta_V=1$, $n_{c,0}=1.6$, $\Delta E=5.5$, with $B = 0$ for (a) and $T=0.005$ for (b). Note the the ground state mixed valency for these parameters is $\langle b \rangle^2 \approx 0.3$ at $B=0$, and the real materials likely have significantly smaller zero field moments.
 \label{fig:magneticmoments}}
 \end{figure}

Broken time-reversal symmetry manifests as magnetic moments for both the conduction electrons, $\vec{m}_c$, and the excited doublet, $\vec{m}_{b}$. These are strictly parallel to the hastatic spinor $\hat b$, and are calculated as,
\begin{align}
m_c &= -\left.\frac{\partial \mathcal{F}}{\partial B_c}\right|_{B_c \rightarrow 0}; & m_b &= -\mu_B g_L \langle J_z \rangle_{\Gamma_7} |b|^2
\end{align}
where $B_c$ is conjugate to $m_c$, coupling only to the conduction electrons. Both magnetizations turn on linearly below $T_K$, as shown in Fig.~ \ref{fig:magneticmoments}(a), which follows from  the BCS-like temperature dependence of $\langle b \rangle$, but is also seen to persist in the Landau theory for small moments and finite $B$.  The total magnitude is small, $O(T_K/D)$, with $D$ the conduction electron bandwidth. 
While the FH moments are quite small in zero field, they will grow fairly quickly in finite fields (Fig.~\ref{fig:magneticmoments}(b)). The most straightforward way to positively identify FH order over the competing quadrupolar orders is to examine the field-dependence of the magnetic moments -- in particular, their direction.  FH moments will always be pinned to the field, and so all moments will align with the external field; by contrast, FQ order induces magnetic moments in field with a significant perpendicular component for some field directions \cite{Ito2011,Sato2012,Taniguchi2016}.  Similarly, AFQ order generically induces FQ order and will have the same field dependence of the uniform moments.  As quadrupolar order is difficult to detect directly, due to its weak coupling to the lattice \cite{Devishvili2008}, measuring the in-field moments along several directions is essential to distinguish between hastatic and quadrupolar orders.

\subsection{Thermodynamic signatures}
 
Broken time-reversal symmetry is also apparent in the development of a finite magnetostriction in the FH state.  For the hastatic spinor aligned along the $z$ axis, this susceptibility is given by $\chi_{ms} \equiv \partial^2 \mathcal{F}/\partial B_z \partial \epsilon_\mu \neq 0 \
\propto \lambda_\parallel$.  Susceptibilities involving $B$ and strain derivatives along $x,y$ vanish, so we do expect a small zero field volume magnetostriction that increases with decreasing temperature in the FH state. A zero field magnetostriction has been observed in PrV$_2$Al$_{20}$ \cite{Patri2018} preferentially along the $[111]$ direction, as one would expect for octupolar order; the hastatic magnetostriction would be relatively independent of the direction of field, as long as field and strain components are aligned. As we believe FH might explain the heavy Fermi liquids in Pr(Ir,Rh)$_2$Zn$_{20}$ at \emph{finite} fields, this signature is not practical, as the transition will be smeared out. 

\begin{figure}[ht]
 \includegraphics[scale=0.65]{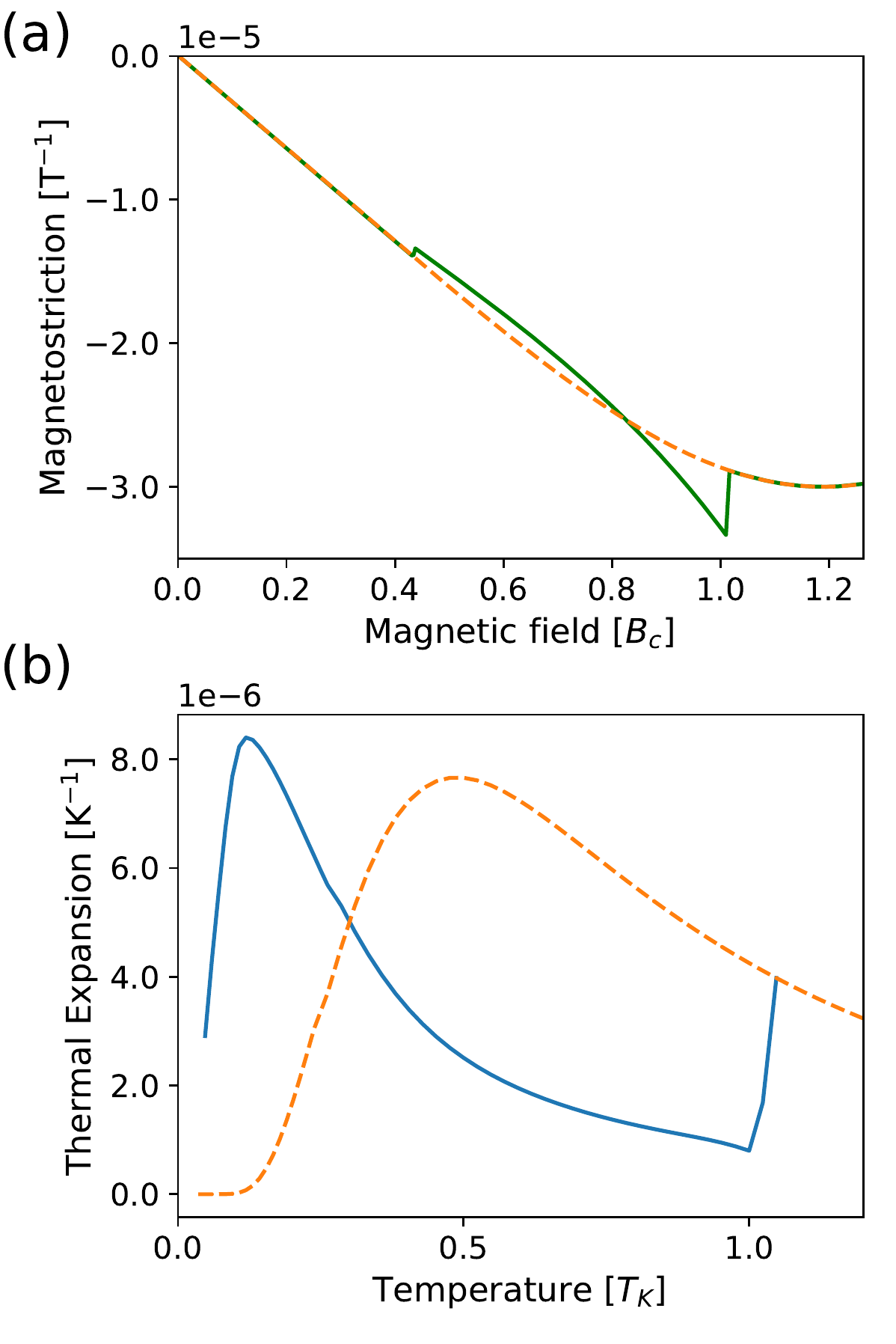}%
 \caption{Mean-field calculations of magnetostriction and thermal expansion. (a) Magnetostriction versus field for the self-consistent mean-field solution (green) and for the paraquadrupolar state (dashed orange); this calculation was done at a relatively large fraction of $T/T_K(B)$. Between $B_z \approx 0.4 B_c$  and $B_z=B_c$ the system is FH. There are jumps in $\lambda_{\parallel}$ when entering the phase, which show a clear increase with increasing field; these are expected to be smeared out by $1/N$ corrections. (b) Thermal expansion versus temperature for the FH (blue) and paraquadrupolar (dashed orange) phases at intermediate fields.  The jump at $T=T_K$ is a signature of the onset of FH order, while the narrow peak at low $T$ contrasts with the broader one of the PQ state. Again, the jump is expected to be smeared out by $1/N$ corrections. Parameters for (a): $t=1$, $\eta_c=1$, $V=0.9$, $\eta_V=1$, $\Delta=20$, $n_{c,0}=1.6$, $T=0.168$; (b):  $t=1$, $\eta_c=1$, $V=0.8$, $\eta_V=1$, $\Delta=10$, $n_{c,0}=1.9$, $B_z=0.8$.  The vertical axes of (a) and (b) are scaled such that the minimum value of the calculated PQ magnetostriction in (a) roughly matches the minimum experimentally determined value at the lowest measured temperature \cite{Worl2018}.
 \label{fig:MSandTE}}
 \end{figure}

The magnetostriction and thermal expansion expected for FH order can also be calculated more generally within the microscopic mean-field model. While we expect the behavior near the transition will be modified as indicated by the Landau theory, the microscopic calculation allows us to access the behavior away from the phase transition.  Fig.~\ref{fig:MSandTE}(a) shows the magnetostriction at fixed temperature $T=1.2T_{K,B=0}$ as a function of field using the parameters of Fig.~\ref{fig:PD}(b). The self-consistently calculated result exhibits jumps at the transitions into and out of the FH phase, but otherwise mostly follows the single ion physics. Fig.~\ref{fig:MSandTE}(b) compares the thermal expansion calculated in the FH and PQ phases (using a different set of parameters than the magnetostriction calculation).  As the temperature is lowered, the thermal expansion exhibits a sharp downward jump upon entering the FH phase, followed by a superlinear rise.  This jump will again be somewhat smeared out, looking similar to the downturns seen in the Landau theory.  Similar behavior has been observed in experiment \cite{Worl2018}, with a dip followed by a steep rise.  The peak in $\alpha_\parallel$ is much sharper and shifted to low temperatures compare to the corresponding result for the PQ phase, which can explain why the experiments have only measured an increasing $\alpha_\parallel$ as the temperature is reduced, having not reached low enough temperatures to see the inevitable downturn.  Note that the valence change associated with hastatic order will also have a small contribution to the volume magnetostriction $\lambda = \lambda_{\parallel} + 2 \lambda_{\perp}$, in addition to the symmetry breaking contribution discussed above.  Recent magnetostriction measurements on PrIr$_2$Zn$_{20}$ suggest relatively small changes in valence as a function of field, in the heavy Fermi liquid region \cite{Worl2018}, consistent with the relatively flat $T_K$ seen in Fig.~\ref{fig:PD}(c).

The magnetic field phase diagrams of PrT$_2$Zn$_{20}$ (T=Ir,Rh), with their intermediate field heavy Fermi liquid regions, are consistent with our model.  PrIr$_2$Zn$_{20}$ orders antiferroquadrupolarly (with $O_2^2$-type moments) for $B = 0$ \cite{Iwasa2017}, but has a finite field region between 4-5T for $B || [100]$ with enhanced $C/T$ and $A$ \cite{Onimaru2016b}, while PrRh$_2$Zn$_{20}$ has a similar heavy Fermi liquid region between 3.5-6.7 T for $B || [100]$ \cite{Yoshida2017}.  Note that an earlier review suggested that these regions could be composite order \cite{Onimaru2016a}; here we propose specifically that these are FH, justified within a microscopic model with concrete predictions. Measuring the magnetic moments for multiple field directions is the best way to differentiate FH and quadrupolar orders. FH order will additionally exhibit a half-hybridization gap in optical conductivity or STM measurements, and Raman measurements of the symmetry breaking hybridization gap are another intriguing possibility. PrV$_2$Al$_{20}$ also exhibits an intermediate in-field region \cite{Shimura2015}, but more experiments are needed.  Other $\Gamma_3$ materials like PrInAg$_2$\cite{Yatskar1996} and PrPb$_3$\cite{Sato2010}, with its high field phases also merit further study.

\section{Conclusions \label{sec:conclusion}}

To summarize, we have investigated ferrohastatic order in cubic systems via a realistic two-channel Anderson lattice model, in combination with a phenomenological Landau theory that accounts for the effect of fluctuations. The development of a heavy Fermi liquid necessarily breaks channel symmetry, including time-reversal and spin rotation symmetries. For FH order, this heavy Fermi liquid includes spin-textured dispersions, symmetry-breaking hybridization gaps, and small magnetic moments for both the conduction electrons and excited $f$-states.  Several materials may realize FH order in finite magnetic field [Pr$T_2$Zn$_{20}$ (T=Ir, Rh)], and it is also a possible candidate for PrTi$_2$Al$_{20}$ once the FQ order is suppressed under pressure.

The nature of two-channel Kondo lattice superconductivity is an open question; thus far research has focused on composite pairing \cite{Flint2008,Hoshino2014,Kusunose2016}.  Quadrupolarly mediated superconductivity arising out of the FH state leads to Cooper pairs that are orbital singlets and spin triplets, at least in the resonating valence bond limit where the orbital singlets first form among the spinless $f$-``electrons'', and are transmitted to the spinful conduction electrons via the hastatic spinor \cite{Andrei1989}. The resulting triplet state is reminiscent of the $A_1$ phase of He-3, due to the asymmetry between $\uparrow \uparrow$ and $\downarrow \downarrow$ pairs\cite{Vollhardt2013}.  Further exploration of AFH order and superconductivity in the hastatic state is left for future work.  

We thank Piers Coleman, Premala Chandra, Peter Orth, and Arun Paramekanti for helpful discussions.  This work was supported by the U.S. Department of Energy, Office of Science, Basic Energy Sciences, under Award No. DE-SC0015891.  R.F.  acknowledges the hospitality of the Aspen Center for Physics, supported by National Science Foundation Grant No. PHYS-1066293.


\appendix

\section{Spin-orbit-coupled Hybridization \label{sec:slaterkoster}}

Here we give further details regarding the spin-orbit coupled hybridization in our model for FH order.  As mentioned in the main text, the valence fluctuation term 
\begin{align}
H_{VF} = V \sum_{j\mu\alpha} \Big[ \tilde{\mu} |j \Gamma_3, \alpha \rangle \langle j \Gamma_7, -\mu | \psi_{j, \Gamma_8 \mu \alpha} + H.c. \Big]
\end{align}
involves the creation or annihilation of conduction electrons projected onto the $\Gamma_8$ symmetry channel of the localized $f$ electrons.  Explicitly, we consider the overlap of a $d$-band conduction electron of $e_g$ symmetry with the $\Gamma_8$ $f$-electron orbital.  Using angular momentum eigenstates $| l, m_l, s, m_s \rangle$ for the $l=2$ $d$ electrons, the $e_g$ states are 
\begin{align}
|e_g,\sigma, + \rangle &= |2, 0, 1/2, \sigma \rangle \\
|e_g,\sigma, - \rangle &= \frac{1}{\sqrt{2}}(|2, 2, 1/2, \sigma \rangle+|2, -2, 1/2, \sigma \rangle ) 
\end{align}
with $\sigma = \pm \frac{1}{2} = \uparrow,\downarrow$.  On the other hand, the $J=5/2$ $\Gamma_8$ quartet states are expressed using total angular momentum eigenstates $| j, m_j \rangle$:
\begin{align}
|\Gamma_8,  \mu, +  \rangle &= | 5/2, \tilde{\mu} (1/2) \rangle \\
|\Gamma_8, \mu, -  \rangle &= \sqrt{\frac{5}{6}}| 5/2, \tilde{\mu} (5/2) \rangle + \sqrt{\frac{1}{6}}| 5/2, -\tilde{\mu} (3/2) \rangle
\end{align}
with $\tilde{\mu} = \mathrm{sgn} (\mu)$.  Since the $\Gamma_8$ electrons in our model arise from the overlap of the $e_g$ conduction electrons with $f$ electron states, the annihilation operators of the former can be written as
\begin{align}
\psi_{\Gamma_8, j, \mu, \alpha} &= \sum_{j',\sigma,\alpha'} \langle \Gamma_8, j, \mu, \alpha | e_g, j', \sigma, \alpha' \rangle c_{j',\sigma, \alpha'}
\end{align}
Here the conduction electron state sits at a distinct site, $j'$, as generically the overlap between d- and f-electrons at the same site is zero; we assume that the $f$-electron is located at the origin, ${\bf R}_{j} ={\bf 0}$. The wave function overlaps are 
\begin{align}
\langle \Gamma_8, j, \mu, \alpha | e_g, j', \sigma, \alpha' \rangle = \!\int\! \mathrm{d}\vect{r} \langle \Gamma_8, j, \mu, \alpha | \vect{r} \rangle \langle \vect{r} | e_g, j', \sigma, \alpha' \rangle
\end{align}
The $e_g$ wave functions are sums of spherical harmonics via
\begin{align}
\langle \vect{r} | e_g, j', \sigma, \alpha' \rangle =  \sum_m &\langle \vect{r} - {\bf R}_{j'} | 2, m,\frac{1}{2}, \sigma \rangle \notag \\
&\times \langle 2, m, \frac{1}{2}, \sigma | e_g, j, \sigma, \alpha' \rangle \\
\langle \vect{r} -  {\bf R}_{j'} | 2, m, \frac{1}{2}, \sigma \rangle &= Y^m_2(\vect{r} - {\bf R}_{j'}),
\end{align} 
while the spin-orbit-coupled $\Gamma_8$ expressions contain additional Clebsch-Gordan coefficients,
\begin{align}
\langle \Gamma_8, j, \mu, \alpha | \vect{r} \rangle = \sum_m &\langle \Gamma_8, j, \mu, \alpha | j, m \rangle \langle j, m | 3, m - \sigma, \frac{1}{2},\sigma \rangle \notag \\
&\times \langle 3, m - \sigma, \frac{1}{2},\sigma | \vect{r} \rangle, \\
\langle j, m | 3, m - \sigma,\frac{1}{2},\sigma \rangle &= -2\sigma \sqrt{\frac{7/2-2m\sigma}{7}},\\
\langle 3,m-\sigma,\frac{1}{2},\sigma|\vect{r}\rangle &= [Y^m_3(\vect{r})]^*.
\end{align}
Following the Slater-Koster method \cite{Slater1954,Alexandrov2013,Baruselli2014}, we numerically calculate the overlaps of wave functions on neighboring sites and determine how they are related by symmetry.  The following overlaps between neighboring $\Gamma_8$ orbitals at position $(r_x,r_y,r_z)$ and $e_g$ orbitals at $(r_x,r_y,r_z + \delta)$ in the $z$ direction are found to be nonzero and generically distinct, with their proportionality captured by the factor $\eta_v$:
\begin{align}
&\langle e_g, \uparrow, + |  \Gamma_8, \uparrow, + \rangle = -\langle e_g, \downarrow, + |  \Gamma_8, \downarrow, + \rangle = \tilde{V} \\
&\langle e_g, \uparrow, - |  \Gamma_8, \uparrow, - \rangle = -\langle e_g, \downarrow, - |  \Gamma_8, \downarrow, - \rangle = \tilde{V}\eta_v . \end{align}
For $e_g$ orbitals located at $(r_x,r_y,r_z - \delta)$, the signs of each overlap are reversed.  This leads to an odd-parity hybridization term along the $z$ direction after the  Fourier transform to momentum space [basis: $(\uparrow +, \uparrow -, \downarrow +, \downarrow -)$]
\begin{align}
H_{e_g - \Gamma_8}^z = i \tilde{V}
\begin{pmatrix}
-2 s_z & 0 & 0 & 0 \\
0 & - 2 \eta_v s_z & 0 & 0\\ 
0 & 0 & 2 s_z & 0\\ 
0 & 0 & 0 & 2 \eta_v s_z
\end{pmatrix}
\end{align}
with $s_z = \sin(k_z)$ Here $\eta_v$ tunes the relative overlap integrals between $e_g$ and $\Gamma_8$.  The 3D hybridization term with cubic symmetry is then obtained from the 1D $H_{e_g - \Gamma_8}^z$ by applying $2\pi/3$ rotations around a cubic body diagonal, transforming $\hat{z} \rightarrow \hat{x} \rightarrow \hat{y} \rightarrow \hat{z}$.
\begin{align}
\mathcal{R}_{2\pi/3} = e^{-i \pi/4} \frac{1}{2 \sqrt{2}} \begin{pmatrix}
-1 & \sqrt{3} & -i & i\sqrt{3} \\
-\sqrt{3} & -1  & -i\sqrt{3} & -i \\
1 & -\sqrt{3} & -i & i\sqrt{3} \\
\sqrt{3} & 1 & -i\sqrt{3} & -i
\end{pmatrix}
\end{align}
Then $H_{3D} = H_{e_g - \Gamma_8}^z + \mathcal{R}_{2\pi/3} H_{e_g - \Gamma_8}^x (\mathcal{R}_{2\pi/3})^\dagger + (\mathcal{R}_{2\pi/3})^2 H_{e_g - \Gamma_8}^y [(\mathcal{R}_{2\pi/3})^\dagger]^2$, leading to the form factor $\Phi^{\sigma \alpha'}_{\mu \alpha}(\vect{k})$ expressing the $\Gamma_8$ creation operator in terms of overlaps with $e_g$ conduction electrons
\begin{align}
&\psi_{\Gamma_8, j, \mu \alpha} = \sum_{\vect{k} \sigma \alpha'} \mathrm{e}^{i \vect{k} \cdot \vect{R}_j} \Phi^{\sigma \alpha'}_{\mu \alpha}(\vect{k})
c_{\vect{k} \sigma \alpha'} \\
&\Phi^{\sigma \alpha'}_{\mu \alpha}(\vect{k}) 
= \begin{pmatrix}
\hat{A} & \hat{B} \\
\hat{B}' & -\hat{A}
\end{pmatrix} \\
&\hat{A} = \begin{pmatrix}
-2i s_z & 0  \\
0 & -2i \eta_v s_z
\end{pmatrix}\\
&\hat{B} = \begin{pmatrix}
\frac{i}{2}(1+3\eta_v)s_+ & -\frac{i\sqrt{3}}{2}(-1+\eta_v)s_- \\
 -\frac{i\sqrt{3}}{2}(-1+\eta_v)s_- & \frac{i}{2}(3+\eta_v)s_+
\end{pmatrix}\\
&\hat{B}' = \hat{B} \, (s_+ \leftrightarrow s_-)
\end{align}

where $s_{\pm} \equiv \sin(k_x) \pm i\sin(k_y)$ and $s_z \equiv \sin(k_z)$.  $\Phi$ possesses an overall amplitude, $\tilde V$ that we set to one, as its effect is already captured by the overall strength of hybridization $V$.  Here the matrix is written in the basis $(\uparrow +, \uparrow -, \downarrow +, \downarrow -)$ for $\sigma=\uparrow,\downarrow$ and $\alpha=+,-$.  Similar techniques are used to obtain the conduction electron dispersion, $\epsilon_{\vect{k}}$ used in the main text by treating the $e_g$--$e_g$ hoppings \cite{Alexandrov2013,Baruselli2014} (see eq. \ref{eq:Hd}).

\section{Ferrohastatic hybridization gaps\label{sec:gaps}}

As discussed in the main text, cubic FH order typically possesses a ``half''-hybridization gap, realized by both symmetry-breaking and non-symmetry-breaking hybridization gaps.  As the Hamiltonian cannot generically be diagonalized analytically, analyzing the gaps is complicated.  We first discuss the general case, where we keep the direction of the hastatic spinor general. Here, we examine the full $f$-electron Green's function, where the symmetry-breaking terms of the $f$-electron self-energy can be isolated; these allow us to clearly discuss the terms entering into the dispersion. Next, we discuss a special case, where some hybridization matrix elements vanish from the dispersion and the problem is analytically tractable. Finally, we discuss how the half-hybridization gap affects the density of states.

\subsection{$f$-electron self-energy}

The heavy quasiparticle band structure is given by the solutions of $\det ( \omega - H_{\vect{k}}' ) = 0$, which do not generically have a closed form.  Still, one may gain insight into the symmetry-breaking properties of FH order by factorizing the determinant as 
\begin{align}
\det ( i\omega_n - H_{\vect{k}}' ) & = \det [g^c(\vect{k},i\omega_n)]^{-1} \cr
& \!\! \times \det \left[ i\omega_n \alpha_0 - \Sigma_f (\vect{k},i\omega_n) \right]
\end{align} 
and examining the $f$-electron self energy, $\Sigma_f (\vect{k},i\omega_n) = \mathcal{V}_{\vect{k}}^\dagger g^c(\vect{k},i\omega_n) \mathcal{V}_{\vect{k}}$.  In particular, we may determine which components of $\Sigma_f$ break symmetries of the underlying cubic lattice, and how these couple to the $SU(2)$ symmetry breaking of the hastatic spinor.  The full hybridized $f$-electron Green's function can be obtained from 
\begin{align}
[\mathcal{G}^f(\vect{k},i\omega_n)]^{-1} = i\omega_n\alpha_0 - \mathcal{V}_{\vect{k}}^\dagger g^c(\vect{k},i\omega_n) \mathcal{V}_{\vect{k}}
\end{align}
where the generic hybridization term is a $4 \times 2$ matrix is 
\begin{align}
\mathcal{V}_{\vect{k}} = V\sum_{\mu} \tilde{\mu} b_{-\mu} \Phi^{\sigma \alpha'}_{\mu \alpha}(\vect{k}) 
\end{align}
and the bare unhybridized conduction electron Green's function is obtained from
\begin{align}
[g^c(\vect{k},i\omega_n)]^{-1} =  \sigma_0 \otimes [ (i\omega_n -\psi_{00}) \alpha_0 - \psi_{01} \alpha_1 - \psi_{03} \alpha_3]
\end{align}
with coefficients given by
\begin{align}
\psi_{00} &= \frac{1}{4}\Tr [\mathcal{H}_c \sigma_0 \otimes \alpha_0] = \mu  - t(1+\eta_c)(c_x + c_y + c_z)\\
\psi_{01} &= \frac{1}{4}\Tr [\mathcal{H}_c \sigma_0 \otimes \alpha_1] = \frac{\sqrt{3}}{2} t(\eta_c-1)(c_x - c_y)\\
\psi_{03} &= \frac{1}{4}\Tr [\mathcal{H}_c \sigma_0 \otimes \alpha_3] = \frac{t}{2}(1-\eta_c)(c_x+c_y-2c_z).
\end{align}
Again, we use the Pauli matrices, $\sigma_\lambda$ and $\alpha_\lambda$ ($\lambda = 0,1,2,3$) to represent the spin and pseudospin degrees of freedom.
Inverting,
\begin{align}
g^c(\vect{k},i\omega_n) &=  \left( \frac{1}{(i \omega_n - \psi_{00})^2 - \psi_{01}^2-\psi_{03}^2} \right) \times \notag \\
&\sigma_0 \otimes [ (i\omega_n -\psi_{00}) \alpha_0 + \psi_{01} \alpha_1 + \psi_{03} \alpha_3]
\end{align}
Thus we find three non-zero components $\sigma_0 \otimes \alpha_i$ ($i=0,1,3$) for $g^c(\vect{k},i\omega_n)$, and hence also for the $f$-electron self energy $\Sigma_f (\vect{k},i\omega_n)$, which has the  matrix structure
\begin{align}
\mathcal{V}_{\vect{k}}^\dagger \sigma_0 \otimes \alpha_i \mathcal{V}_{\vect{k}} &= \sum_{j}V_{2,ij}(\vect{k}) \alpha_j
\end{align}
with
\begin{widetext}
\begin{align}
V_{2}^{ij}(\vect{k}) &= \frac{1}{2}\Tr [ \mathcal{V}_{\vect{k}}^\dagger \sigma_0 \otimes \alpha_i \mathcal{V}_{\vect{k}}\alpha_j ] = \begin{cases} \frac{V}{4} \Tr [\Phi(\vect{k})^\dagger \sigma_0 \otimes \alpha_i \Phi(\vect{k}) \sigma_0 \otimes \alpha_j] \Tr[ \hat{b}^\dagger \mu_0 \hat{b}],& \;\; j = 0,1,3 \\
-\frac{V}{4} \sum_k \Tr [\Phi(\vect{k})^\dagger \sigma_0 \otimes \alpha_i \Phi(\vect{k}) \sigma_k \otimes \alpha_2] \Tr[ \hat{b}^\dagger \mu_k \hat{b}],& \;\; j = 2 \end{cases}
\end{align}
\end{widetext}
where $\mu_k$ is one of the Pauli matrices representing the excited Kramers doublet pseudospin. These self-energy terms form a non-trivial matrix in $\alpha$ space, and so the full dispersion will contain not only these terms, but quartic traces of the form,
\begin{equation}
V_{4}^{ij}(\vect{k})= \Tr[ \mathcal{V}_{\vect{k}}^\dagger \sigma_0 \otimes \alpha_i \mathcal{V}_{\vect{k}} \sigma_k \otimes \alpha_j \mathcal{V}_{\vect{k}}^\dagger \sigma_0 \otimes \alpha_i \mathcal{V}_{\vect{k}} \sigma_k \otimes \alpha_j], 
\end{equation}
where we keep only the nonzero terms.
We note that out of the above terms, only $V_{2}^{00}(\vect{k})$ explicitly preserves the cubic symmetry of the underlying lattice for generic values of $\eta_v$.  However, only the terms that also break $SU(2)$ symmetry lead to cubic symmetry breaking in the full dispersion.  The other terms should be considered similarly to the $\eta_c \neq 1$ terms in the conduction electron dispersion: they lead to band splitting, but the overall dispersion satisfies cubic symmetry.  We have checked this explicitly by setting the $b^\dagger \vec{\mu} b$ terms to zero, while keeping the $b^\dagger b$ terms, and have found that cubic symmetry is always preserved.  In order to analyze the symmetry-breaking nature of the hybridization gaps that depend on $b^\dagger \vec{\mu} b$, we now turn to an analytically tractable special case.

\subsection{Analytic expression for $\eta_c = 1$, $\eta_v=\!\sfrac{-1}{3}$, $\hat{b} = (1,0)^T$}

In this special case, the mean-field Hamiltonian may then be diagonalized to obtain two degenerate unhybridized conduction electron bands and the following four hybridized bands:
\begin{align}
E_{\vect{k}} = \! \left[\! \frac{(E_{c\vect{k}}  + \lambda)}{2}\! \pm \!  \sqrt{\frac{[ (E_{c\vect{k}} -\lambda)^2 + V_{1\vect{k}}^2 ]}{4} \pm \frac{16 \sqrt{\gamma_{\vect{k}} + \delta_{\vect{k}}}}{9}} \right]
\end{align}
where we define
\begin{align}
E_{c\vect{k}} &= -2t(c_x+c_y+c_z)+\mu \\
|V_{1\vect{k}}|^2 &= 4 V_{2}^{00}(\vect{k}) = \frac{80}{9} V^2b^2 (s_x^2 + s_y^2 + s_z^2) \\
\gamma_{\vect{k}} &= V^4 b^4(s_x^4 + s_y^4 + s_z^4)\\
\delta_{\vect{k}} &= V^4 b^4 (2s_x^2s_y^2-(s_x^2+s_y^2)s_z^2)
\end{align}
Here, $\gamma_{\vect{k}}$ has the full ($\Gamma_1$) symmetry of the lattice, as does $V_{1\vect{k}}$. $\delta_{\vect{k}}$ breaks the cubic symmetry, and has the symmetry of $|\Gamma_3,+\rangle$, mixing both $g$-wave ($7[2z^4-x^4-y^4]-6[3z^2-r^2]r^2$) and $d$-wave ($3z^2-r^2$) components of the same symmetry; it is plotted in Fig. 2(b) of the main text. $\delta_{\vect{k}}$ is the only term that depends on $b^\dagger \vec{\mu} b$, and is proportional to $(b^\dagger \mu_3 b)^2$.  It is written in terms of the above traces as,
\begin{equation}
\delta_{\vect{k}} = 4 \left(V_{4}^{33} +V_{4}^{11} -V_{4}^{00} -V_{4}^{22}\right) -4 (V_{2}^{32})^2,
\end{equation}
where we have suppressed the $\vect{k}$ dependence on the right hand side.
Note that individually, each $V_4$ or $V_2^2$ is positive definite, and each have different symmetries that are not $|\Gamma_3,+\rangle$; it is only the combination of these that gives the nodal gap.  Rotation of the hastatic spinor to $\hat x$ or $\hat y$ maintains the same shape of the symmetry-breaking gap component, but with a rotated quantization axis.  Thus for $\hat{b}$ along the $\hat{x}$ axis ($ = (1,1)^T/\sqrt{2}$), the analogous component has the form of Fig. 2(b) of the main text, but now oriented along the $\hat{x}$ axis.  Essentially, rotating the hybridization spinor away from $\hat z$ mixes the $\Gamma_3 \pm$ states.


%

\end{document}